\documentclass[12pt,preprint]{aastex}

\shorttitle{New Brown Dwarfs in USco}
\shortauthors{}

\begin{document}

\title{A Large-Area Search for Low Mass Objects in Upper Scorpius I: The 
Photometric Campaign and New Brown Dwarfs}

\author{Catherine L. Slesnick, John M. Carpenter, 
\& Lynne
A. Hillenbrand}

\affil{Dept.\ of Astronomy, MC105-24, California Institute of
Technology,Pasadena, CA 91125}

\email{cls@astro.caltech.edu, jmc@astro.caltech.edu, lah@astro.caltech.edu}

\begin{abstract}

We present a wide-field photometric survey covering $\sim$200 deg$^2$
toward the Upper Scorpius OB association.
Data taken in the $R$ and $I$ bands with the Quest-2 camera on the Palomar 48-inch
telescope were combined with the 2MASS $JHK_S$ survey and used to select
candidate pre-main sequence stars.  Follow-up spectroscopy with the 
Palomar 200-inch telescope of 62 candidate late-type members 
identified 43 stars that have surface gravity signatures consistent with association membership.
From the optical/near-infrared photometry and derived spectral types
we construct an HR diagram for the new members and find 30 likely new brown dwarfs, nearly doubling the known substellar population of the Upper Scorpius OB association.
Continuation of our spectroscopic campaign should reveal hundreds on new stellar and substellar members.
 
\end{abstract}

\section{Introduction}

The Upper Scorpius OB Association (USco) is the closest (145 pc; de Zeeuw
et al. 1999) young OB association to the Sun with  
114 known high mass Hipparcos stars.
At an age of $\sim$5 Myr (Preibisch et al. 2002), this cluster is at an intermediate age between very young star forming regions and older open clusters where samples are sparser and
studies of processes such as circumstellar disk dissipation are critical.
Recent mid-infrared work by Mamajek et al. (2004) and Silverstone et al. (2006) 
indicates that by $\sim$10 Myr dust 
is removed from the inner few AU of circumstellar disks for $\gtrsim$85\% of stars,
whereas $\sim$80\% of young 1 Myr stars in Taurus still retain their disks
at these radii (Kenyon \& Hartmann 1995; Skrutskie et al. 1990).
This evolution in circumstellar material corresponds to the stage when planets are thought to be forming.  Meteoritic evidence suggests the timescale for 
dissipation of our own solar system's nebula
was on order of $\sim$10$^7$ yr (Podosek \& Cassen 1994).
Further, discovery of $^{60}$Fe in meteorites argues that 
short-lived radionuclides were injected into the solar system's early 
protoplanetary disk from the explosion of a nearby supernovae (Desch \& Ouellette 
2005; Tachibana \& Huss 2003).  The evidence strongly suggests that our solar sytem 
was formed in an OB association similar to USco.  Therefore, if we are to understand our own earth's 
origins, we must study 
the evolution of
OB association members during planet-building stages.

A major difficulty faced by studies of the USco region is that
 the Hipparcos members alone span $>$130 deg$^2$ on the sky. Obtaining a 
complete census of the association's low mass population is thus a formidable  
task as one must identify faint objects over a very large spatial region.  While there exist several techniques to identify young stars not associated with molecular gas, many of them are also accretion diagnostics.  
For example, a common method is to search for strong H$\alpha$ emission (Ardila et al. 2000)
produced in outflow or accretion flows (see Section 3.4), or near-infrared excess emission associated with warm inner accretion disks.  
While accretion can terminate over a wide range in age (1--10 Myr), the median
lifetime of optically thick accretion disks is closer to 2--3 Myr (Haisch et al. 2001; Hillenbrand 2006).  Therefore, surveys to look for accretion signatures alone will not garner a full census of a 5 Myr association.

Enhanced chromospheric and coronal activity can last well beyond accretion timescales.  This activity is linked with X-ray emission
(Ku 
\& Chanan 1979; Feigelson \& DeCampli 1981) though the exact cause of this
phenomenon is still not fully understood (Preibisch et al. 2005).
Many previous large-scale efforts in USco have successfully utilized Einstein data (Walter et al. 1994) or the ROSAT All Sky Survey (RASS; Preibisch et al. 1998,1999) to identify hundreds of low and intermediate mass association members.
However, neither the Einstein observations nor the RASS were sensitive enough
to detect faint X-ray emission from the lowest mass stars and brown dwarfs.

Recently, deep, multicolor imaging surveys combined with spectroscopic follow-up 
have proved successful in identifying both the youngest classical T 
Tauri-type objects and more evolved very low mass stars and
 brown dwarfs in a variety of young regions.  Young PMS objects still undergoing contraction 
towards the main sequence are redder and more 
luminous
than their main sequence counterparts.
Spectroscopic follow-up observations allow assessment of surface gravity diagnostics which can be used to distinguish young PMS stars from reddened field dwarfs
and background giants.
Previous imaging and spectroscopic surveys in USco include work by Preibisch et al. (2001,2002) who selected candidate association members based on optical magnitudes and 
colors obtained from photographic plates in the United Kingdon Schmidt Telescope
survey fields.  Their spectroscopic survey of 700 candidates over 9 deg$^2$ using the 2dF
multifiber spectrograph yielded 166 new PMS stars based on the presence of lithium in their spectra.  Mart\'{i}n et al. (2004) selected candidate young objects from the DENIS $I,J$ survey and obtained spectra of 36
targets.  Of these 28 were confirmed to be new low mass association members based on surface gravity diagnostics.  
Ardila et al. (2000) used 
$R,I,Z$ photographic photometry to identify candidate members within an
80'$\times$80' area of the association.  Spectroscopic data were obtained for 22 candidates, 20 of which were 
determined to exhibit H$\alpha$ emission indicating possible membership.
Thus far,
over 300 low mass ($M<$0.6 M$_\odot$) members have been identified in USco through X-rays, H$\alpha$ emission, photometry and/or spectroscopy. However most searches have been limited to 
small subregions or bright objects. 
Given the USco  upper
IMF, and assuming the high and low mass objects share the same spatial 
distribution, Preibisch et al. (2002) estimate the entire USco region should 
contain $>$1500
young, low mass objects with $M<$0.6 M$_\odot$, most of which are yet to be discovered.

Building on previous work in this region, we have completed a large-area optical 
photometric survey of $\approx$200 deg$^2$ in and near USco.
We combine these data (section 2) with 2MASS $J,H,K_S$ photometry to select 
candidate young PMS objects.  In section 3 we present newly-obtained
optical spectra for 62 candidate young brown dwarfs and low mass stars.
These spectra allow us to determine spectral type and confirm
membership for photometrically-selected candidates.  
Finally in section 4, we derive an HR-diagram and
discuss noteworthy new members.

\section{Observations}

\subsection{Photometric Monitoring and Data Processing}

Photometric observations were taken with the Quest-2 Camera (Rabinowitz 
et al. 2003) on the 48-inch Samuel Oschin Schmidt Telescope at Palomar 
Observatory.
The Quest-2 Camera is a large-area mosaic of 112 CCDs arranged 
in a grid of 4 columns by 28 rows.  Each CCD has 
600$\times$2400 pixels with a scale of $\sim$0.8''/pixel.
The camera covers a 3.6$^o$$\times$4.6$^o$ field of view.  Taking into account  
gaps between columns and chips the total sky coverage is 9.4 deg$^2$ per pointing. 
Each of the four columns views the sky through a separate filter.

Data were obtained in driftscan format using $U,B,R,I$ filters.  Any given patch of sky is observed 
over the entire 2400-pixel width of four separate CCDs (1 per filter) in one of the 28 
rows.  Charge is continuously read out of each CCD throughout the observation
and the final data product is a strip of uniform width in declination and time-dependent 
length in right ascension.  Three scans, centered at 
$\delta=-15.7^o,-19.5^o,-23.3^o$, were each observed between RA of 15h46m and 
16h36m.
The scan centered at $\delta=-19.5^o$ was observed
 3--4 times per night 
on seven consecutive, photometric nights between 20-26 June 2004.
The other two scans were observed once during this period.
Spatial coverage is shown in Figure~\ref{fig:spat}.  
Because several CCDs have failed since installation due to bad connections or 
faulty chips, our spatial coverage within the survey 
region is not uniform.
Five of 28 CCDs in the $I$ column, five in the $R$ column and four in the $B$ column
are not fully functional.  The CCDs are less sensitive in the $U$-band than anticipated and very few source 
detections were obtained.
As discussed in Section 2.3, we are primarily concerned with selecting faint red
objects as candidate young low mass stars, the majority of which do not have a $B$-band detection. 
We therefore exclude the $U$ and $B$ 
data from the remainder of this discussion.

When operated in driftscan format, the Quest-2 camera generates $\approx$50 GB of 
raw, compressed 
data in a single night.
To efficiently process such a large volume of data, the Quest-2 
Collaboration (Rabinowitz et al. 2003) has developed automated 
data reduction software (Andrews 2003).  
Each of the CCDs is treated as a separate instrument during reductions and 
therefore
we explain the procedure for a single chip.   
The software first performs the basic 
bias subtraction and flat-fielding.
The bias level for a given column of the CCD is computed by median-combining 25 rows 
in the overscan region and subtracted.
Dark subtraction and flat-fielding 
cannot be carried out using the standard techniques of pointed observations.
Instead, skyflat and dark driftscans were taken on the first night of observations.
Each calibration image was first 
divided into 240 600x10 pixel segments and then median-combined using the IRAF
IMSUM task to remove cosmic rays (darks) or stars (skyflats).  The resulting 
600x10 pixel image was then averaged into 
a single row of 600 pixels using the IRAF BLKAVG task.
The averaged dark and flat rows were subtracted and divided respectively from each 
row of data.

After pre-processing, the pipeline generates a point-spread function (PSF) for every frame (600x2000-pixels) of data which it uses to detect sources. It registers detections from all four
filters and generates an astrometric solution from  
the USNO-A2 catalog.  
We have matched detections within a 0.8'' radius from up to 24 different scans in the monitoring region. 
For each source we computed new coordinates by averaging together coordinates
from individual scans and find typical astrometric residuals for individual detections about the mean to be $\pm$0.1''.

The Yale pipeline generates photometry both from a range of apertures and from PSF function fitting.  
We compared PSF and aperture
photometry and found the aperture photometry produced more consistent results
for sources observed in multiple scans (see Section 2.2 for a detailed discussion of photometric reliability).
We therefore 
chose to rely on aperture photometry
despite the potential advantages of PSF fitting for crowded stars.  
The pipeline measures all objects through an aperture of half-width 3.5 pixels (found to 
produce the highest signal to noise photometry for 
a subset of bright, isolated stars). 
To account for zeropoint variations between chips we applied 
the calibrations derived by the Quest-2
instrument group at Yale (Jerke et al. private communication).   These calibrations also 
converted Quest-2 instrumental magnitudes into the Johnson 
photometric system.
 
We 
accounted for small atmospheric transparency changes during the night and between 
nights by applying a photometric offset to 
every scan as a function of RA.   
We selected a subset of calibrator stars from our source catalog that were detected in at 
least 20 out 
of 24 of the $\delta =$-19.5 monitoring scans, 
and had no neighbors within 5''. For each star and filter an uncertainty-weighted average 
magnitude of all detections was calculated along with
the difference between that average magnitude and the magnitude measured
 on each individual scan. 
For every chip we created a catalog of offsets by stepping
through in RA every 25 calibrator stars and calculating a median offset value.    
We applied these offsets of up to 0.25 mag to the entire data set as a function of RA and CCD chip 
by linearly interpolating 
in RA values. 
For the high ($\delta =$-15.7$^o$) and low ($\delta =$-23.3$^o$) declination scans 
which overlap only the top and bottom 0.8 deg of the monitoring scan region, 
we calculated a scan offset as a function of RA using the same procedure outlined 
above 
but averaging together all columns of overlap with the mid-declination scans, 
rather than chip by chip.  We 
find this procedure produced magnitudes consistent with those derived from the 
multiple-scan region at the 7\% ($R$) \& 2\%($I$) levels for objects in the overlap region.
Finally, to account for differences in airmass between the Yale
calibrator scans and USco scans we applied offsets of 0.08 \& 0.05 mag for the
$R,I$ data (Parrao \& Schuster 2003) as derived at San Pedro M\'{a}rtir
and consistant with typical Palomar extinction coefficients.

\subsection{Precision and Accuracy}

In this section we use photometry from the monitoring scans to assess the relative precision of the photometric data.  We find photometric precision to be CCD dependent and give values for best, worst and typical scenarios.  The photometric accuracy is harder to quantify at this time.  Zeropoints for individual CCDs and conversions from Quest-2 to Johnson photometry have not yet been finalized by the Quest-2/Yale instrument team.  To place some constraints on data accuracy we have matched the $R$- and $I$-band photometry to photometry from Ardila et al. (2000) and examine
consistency. 

\begin{figure}
\scalebox{0.7}{\includegraphics{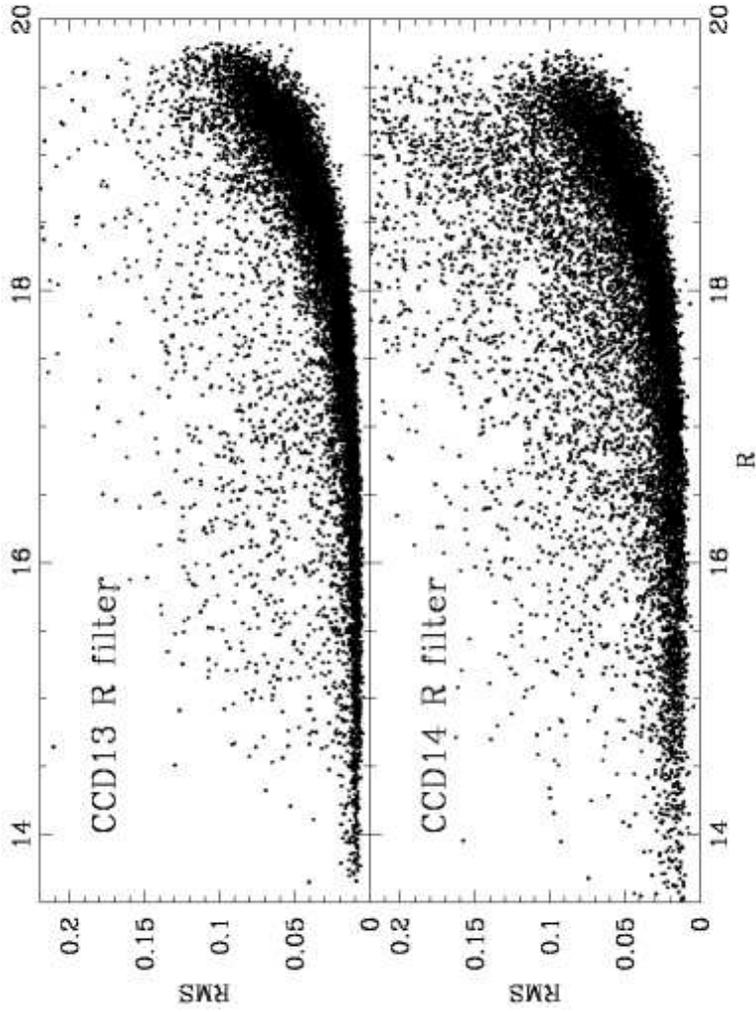}}
\caption{Computed RMS deviations as a function of magnitude for two CCDs.  Each magnitude is the average of 10--24 measurements and the corresponding RMS is a weighted
deviation of all individual measurements about that average.  The top and bottom panels show respectively repeatability plots for a `best' and typical CCD.  See text for
further explanation.}
\label{fig:rms}
\end{figure}

For each source in the multiple scan region, an uncertainty-weighted average 
and the RMS deviation of the individual measurements was calculated.
The 112 CCDs are of varying quality which leads to non-uniform photometric characteristics.  
Figure~\ref{fig:rms} shows the RMS as a function of magnitude for a single CCD for sources detected at least ten times.
The top panel shows data for the CCD in row 13, column 1 ($R$ filter), 
 one
of the most reliable CCDs.  The bottom panel shows data for an adjacent CCD in row 14, column 1, representative of
a typical CCD. 
The best CCDs have 
RMS values from $\sim$0.01 mag for the bright stars and increasing to $\sim$0.05 mag by $R$=19 and $I$=18.5.
A typical CCD has  RMS$\sim$0.03 mag at the bright end to $\sim$0.08 mag at the faint end.  We find six CCDs to be of poor quality with RMS ranging 
from $\sim$0.07--0.15 mag at the bright and faint ends.

\begin{figure}
\epsscale{0.8}
\plotone{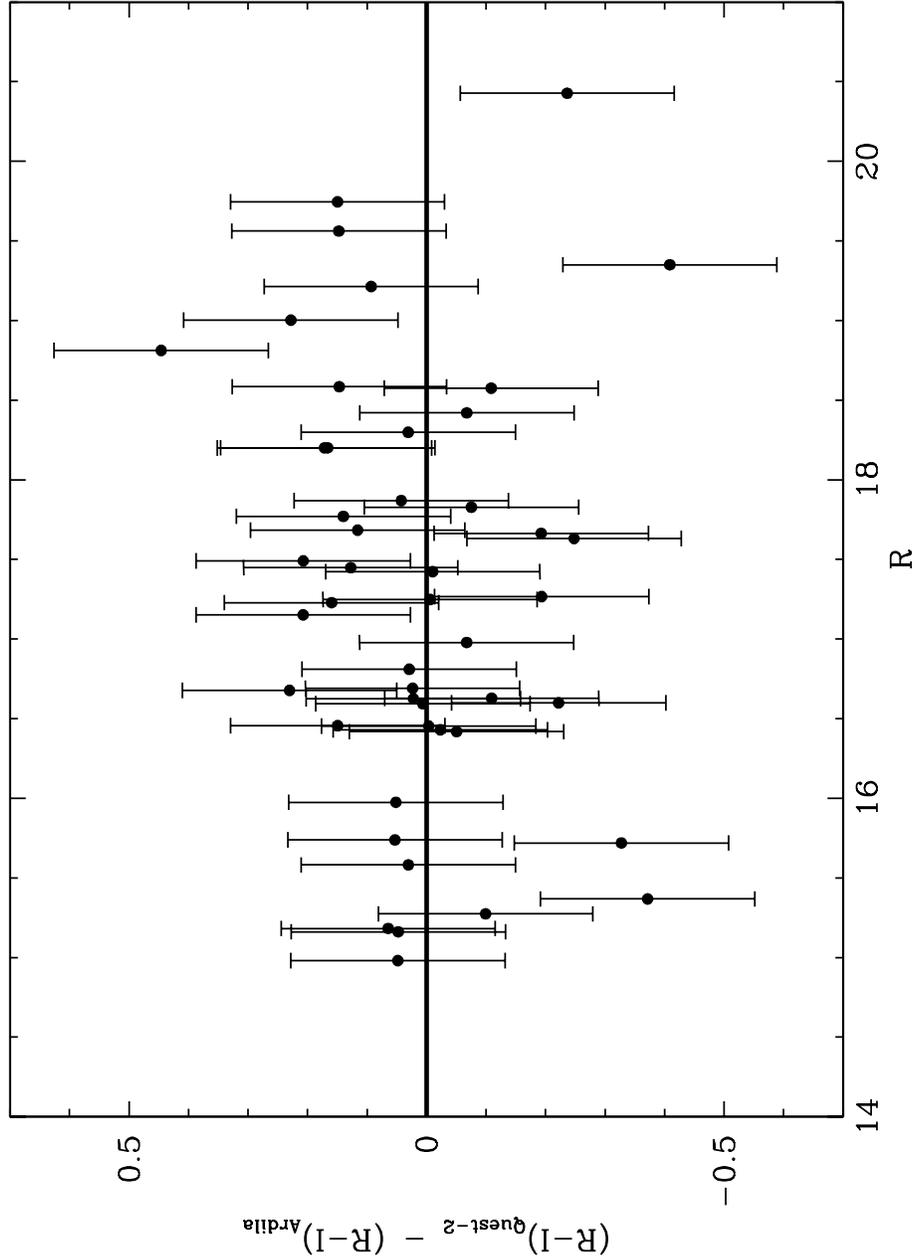}
\caption{Difference between Quest-2 $R-I$ colors and $R-I$ colors from Ardila et al. (2000) (converted to the Johnson system) as a function of Quest-2 $R$ magnitude.  Errorbars reflect root-sum squared uncertainties in both
datasets.}
\label{fig:amb}
\end{figure}

The accuracy of the photometric transformations from the Quest-2 to the Johnson system is harder to quantify since
differences both in zeropoints between chips and in conversion between photometric systems must be accounted for.  The Quest-2 instrument team at Yale has matched Quest-2 observations in the north to Sloan Digital Sky Survey (SDSS)
photometry and derived conversions for each chip from Quest-2 to SDSS (converted to the 
Johnson system; Fukugita et al. 1996).
We have used the current calibrations and 
attempted to quantify the accuracy of the photometry presented
here by comparing our data to that of Ardila et al. (2000) who 
observed $\sim$14 deg$^2$ in USco in the Cousins $R$ and $I$ bands.
Figure~\ref{fig:amb} shows the difference between Quest-2 Johnson $R-I$ colors and those from 
Ardila et al. (2000) converted from the Cousins to the Johnson system (Fernie et al. 1983) as a function of Quest-2 Johnson $R$ magnitude.
Error bars of $\pm$0.18 mag reflect average root-sum squared uncertainties from both data sets: $<\sigma_{R-I}>$=0.14 in the 
Ardila et al. (2000) photometry (private communication) and $<\sigma_{R-I}>$=0.11 for the Quest-2 photometry assuming typical magnitude uncertainties of 0.08 mag for faint stars.  The average difference between the two data sets is 0.01 mag with an RMS 
deviation of 0.17, consistant with uncertainties.

From histograms 
of all detections for the
$R$ and $I$ filters, we find a turnover in number of objects detected at
$R\sim$19 mag and $I\sim$18 mag with a substantial number of detections out to $R\sim$20 and $I\sim$19.  
We believe the precision of our photometry to be $\lesssim$0.08 mag for $\sim$90\% of stars brighter than these values.
Both $R$ and $I$ data saturate for stars brighter than $\sim$13 mag.

\subsection{Candidate Member Selection}

\begin{figure}
\plotone{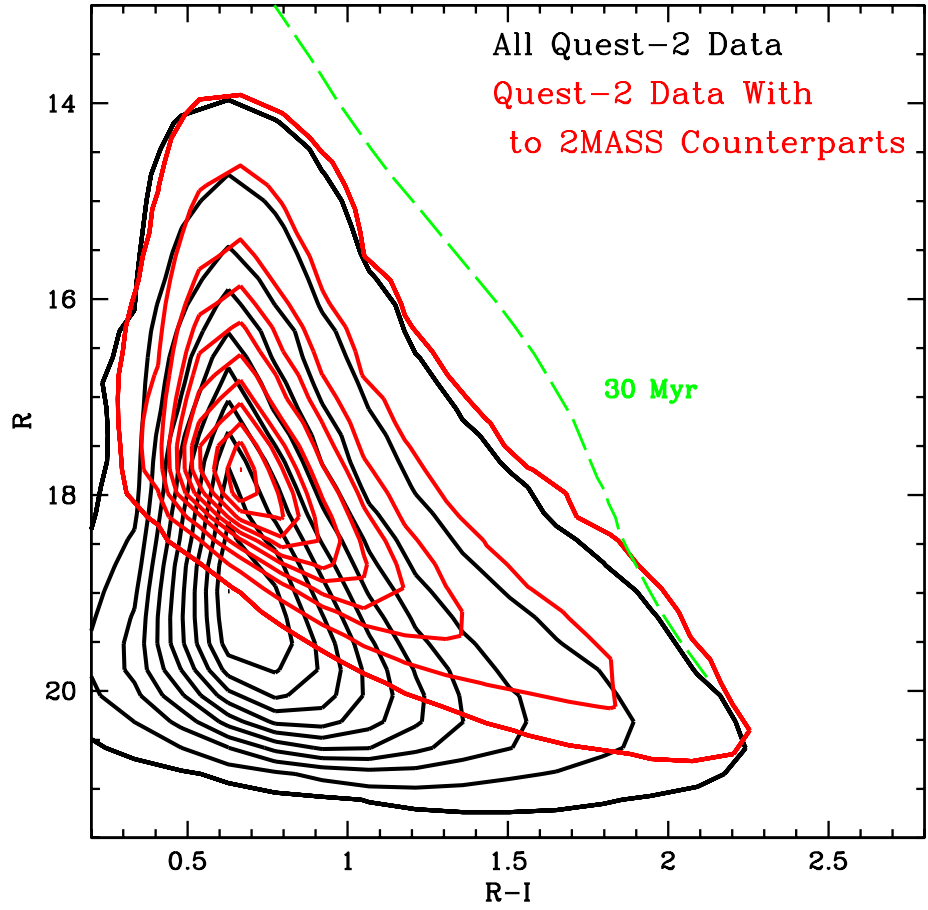}
\caption{Optical color-magnitude diagrams of all Quest-2 sources (black) and those with 2MASS detections (red).  For both black and red contours data are represented at 90\% to
10\% and 3\% of the peak level.  }
\label{fig:2mass}
\end{figure}

The detection algorithm in the Yale data pipeline was written to find
very faint quasars and therefore uses a low detection threshold. 
To produce a reliable catalog of sources,  we first required an object 
to be detected in both the $R$ and $I$ filters.
Further, we matched our 
entire dataset to the 2MASS All Sky Catalog (Cutri et al. 2003) and considered stars
as potential candidates only if they had a 2MASS counterpart 
within 2''.  
This cut biases 
our final catalog against faint blue objects to which 2MASS is not sensitive. 
Of the $\sim$1.5 million objects in our source catalog, $\sim$500,000 have
2MASS counterparts.  However, because we are interested only in objects redder than
a 30 Myr isochrone in an optical color-magnitude diagram (see below for details), 
and bright enough to be observed spectroscopically ($R\lesssim$20 mag),
this bias does not affect candidate selection in practice.
Figure~\ref{fig:2mass} shows contour optical color-magnitude diagrams of all Quest-2 sources (black) and those with 2MASS detections (red).  
As can be seen, the red side of the color-magnitude diagram brighter than $R\sim$20 is not substantially affected by
excluding objects without a 2MASS detection.

\begin{figure}
\plotone{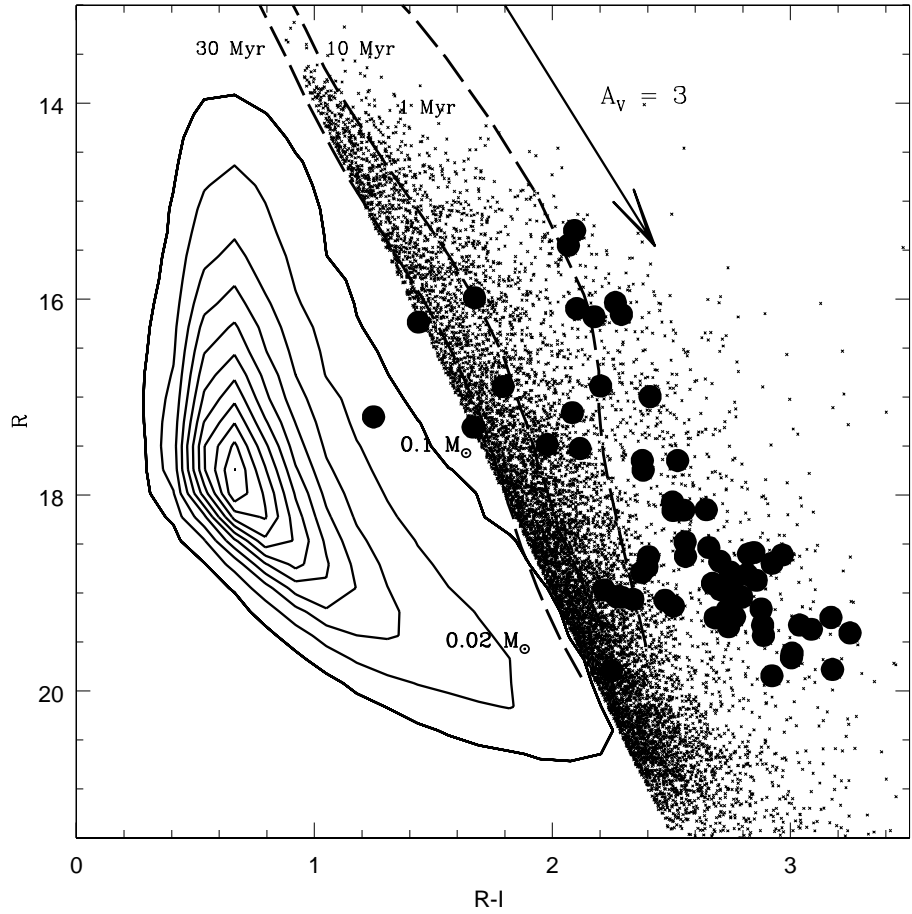}
\caption{$R,R-I$ color-magnitude diagram for all sources with a 2MASS detection, shown with isochrones
from DM97 transformed by us into the $R,R-I$ plane.  Contours are as described in Figure~\ref{fig:2mass}. Objects which appear younger than $\sim$30 Myr are shown as 
discreet points.  Spectroscopic targets presented here are shown as large circles. }
\label{fig:cmd}
\end{figure}

\begin{figure}
\plotone{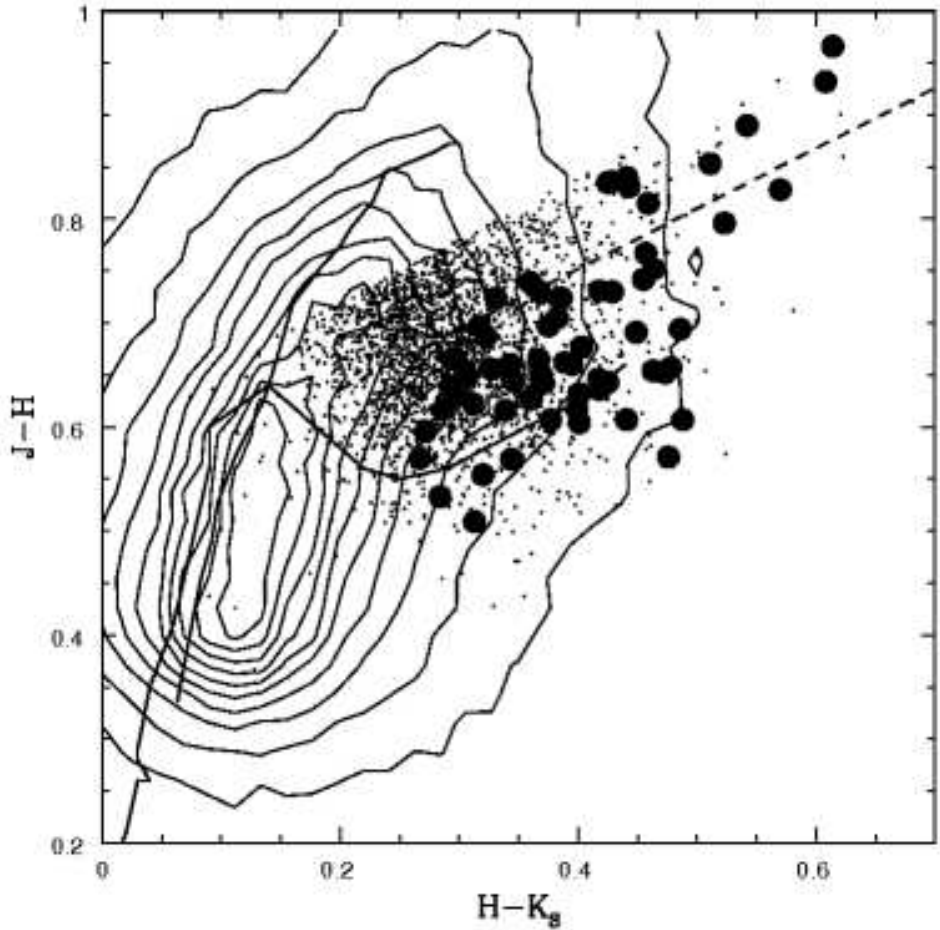}
\caption{2MASS $J-H,H-K_S$ color-color diagram for all data represented as contours at 90\% to
10\% and 3\% of the peak level.
Shown as discreet points are all objects which appear younger than
30 Myr in an optical CMD and do not have colors consistent with those of background
giants.  Solid lines are dwarf zero-age main sequence (O5--M8) and giant loci (G0--M7).  The dashed line represents the classical T Tauri locus as defined by Meyer et al. (1997).  Spectroscopic targets presented here are shown as large circles.  }
\label{fig:ccd}
\end{figure}

\begin{figure}
\plotone{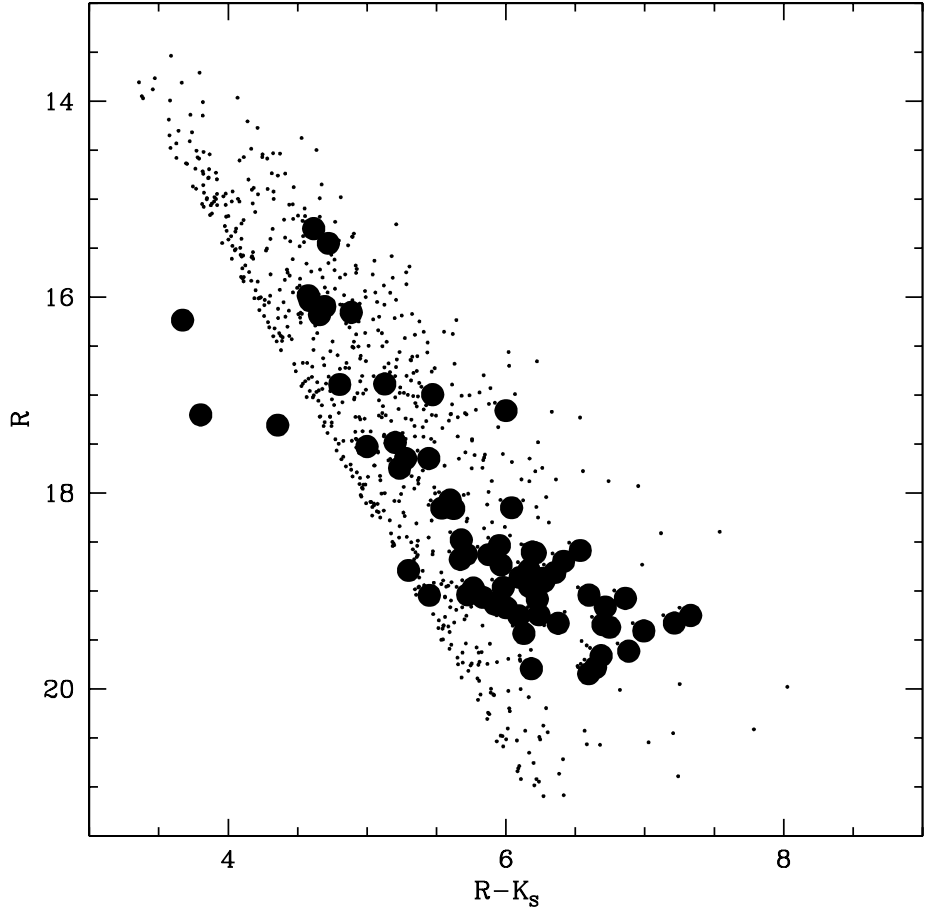}
\caption{$R,R-K$ color-magnitude diagram for the $\sim$1000 objects which meet optical and near-infared selection criteria shown in Figures~\ref{fig:cmd} \& ~\ref{fig:ccd} 
and which appear red in $R-K$. Spectroscopic targets presented here are shown as large circles.  }
\label{fig:cmdrk}
\end{figure}

The final photometric catalog that we use to select candidate association members contains $\sim$500,000 sources with $R, I, J, H$, \& $K_S$ detections, most of which are 
field stars.
However, buried in these $\sim$500,000 sources are several thousand bona fide 
low mass members of USco.
We use this data-set to select candidate PMS stars based on 
an optical color-magnitude diagram indicating 
they could be young USco members.  In Figures~\ref{fig:cmd}--\ref{fig:cmdrk} we present 
color-magnitude and color-color diagrams used in PMS candidate selection.  
For the work presented here, we used three levels of selection criteria.  In each Figure objects which meet the outlined selection criteria $and$ all previous selection criteria are shown as discreet 
points.
In Figure~\ref{fig:cmd} we show
the $R,R-I$ color-magnitude diagram for all sources matched to 2MASS.  
About 60,000 sources with $R < 5\times(R-I)+11$ appear younger than $\sim$30 Myr 
based on theoretical 
isochrones 
(D'Antona \& Mazzitelli 1997 \& 1998\footnote{The 1998 models are a web-only 
correction at
$<$0.2 M$_\odot$ to their original 1997 work.}, hereafter DM97) and could be 
PMS objects. 

2MASS photometry was used to further refine the candidate sample.  
In Figure~\ref{fig:ccd} we show a near-infrared color-color diagram.  Solid lines
represent the dwarf (bottom line) and giant (top line) loci.  The dashed line
represents the classical T Tauri star locus as defined by Meyer et al. (1997).
Beyond this point we considered only objects
which had `good' 2MASS photometry (quality flag `A') in all three $J,H,K_S$ bands.
Any star with $J-H,H-K_S$ colors consistent with those expected for background 
giants ($(J-H) > 0.6\times(H-K_S)+0.6)$) was excluded.  

After applying the above selection criteria we identified $\sim$20,000 candidate new 
young USco members over $\sim$200 deg$^2$.
Due to interstellar extinction and distance effects, optical and near-infrared colors and 
magnitudes alone are not a unique indicator of youth; therefore, it is necessary
to obtain spectroscopic follow-up observations to determine an object's spectral 
type 
and confirm low surface gravity consistent with that of young association members.
  We began a spectroscopic follow-up program with 5 observing nights in June 2005 
using the Double Spectrograph on the Palomar 200-inch Telescope and 5 more in July 
2005 using Hydra on the CTIO 4-m
Telescope.   
With the Palomar spectroscopic observations discussed
here, we chose to target potential new brown dwarfs and very low mass 
stars 
to take full advantage of the high sensitivity of the Double Spectrograph. 
We therefore selected only the reddest candidates satisfying $R < 2.57\times(R-K_S)+13$ in an $R,R-K_S$ 
color magnitude 
diagram (Figure~\ref{fig:cmdrk}) and focused primarily on the 
faintest ($R>$18.5) targets.  With hydra/CTIO we took
spectra of close to 1000 brighter (15$<R<$18.5) candidates,
leading to new higher mass (spectral type G-M) USco members which will be 
discussed in a forthcoming paper.  After applying this last selection criterion
we were left with $\sim$1000 candidates suitable for spectroscopic follow-up with the Double
Spectrograph, $\sim$200 of which  are fainter than $R$=18.5.

\subsection{Optical Spectroscopy}

Moderate resolution spectra of 65 objects (Table 1; 
shown as large circles in Figures~\ref{fig:cmd}--~\ref{fig:cmdrk}) were obtained
during the nights of 2005 June 8-12 using 
the Double Spectrograph on the 200-inch
Palomar telescope.  Of these, 41 were fainter than $R\sim$18.5 corresponding to 
$\sim$20\% of the faint candidates meeting all
of the selection criteria detailed in Section 2.3.  
Because Quest-2 is a new instrument
and software is continuously being updated, the photometric data has been re-reduced several times.  Photometry for three of the spectral targets has changed 
significantly since the spectroscopic observations were taken
such that they would no longer be considered candidates.
As expected, all three were determined to be field dwarfs based on the their
spectra.  
For completeness these three sources are shown in the figures and listed in Table 1, but otherwise are not discussed further.

Data were taken using the red side of the Double Spectrograph which 
has a 1024$\times$1024 
CCD with 24-$\mu$m pixels.  We used a 2.0'' entrance slit, 5500 
\AA$\;$$\;$dichroic and 316 lines/mm grating blazed at 7500 \AA$\;$ which 
gave a wavelength coverage of 6300--8825 \AA$\;$ at resolution R$\sim$1250.  Typical 
exposure times were 600-900 
seconds and we were able to
observe objects as faint as $R\sim$20 in 1800 seconds with SNR$\sim$30--40.  
Spectrophotometric
standard stars (Massey et al. 1988) were observed throughout every night 
for flux calibration.
All sources, including standards, were pre-processed, extracted and flux-calibrated 
using standard IRAF tasks.

\section{Spectral Analysis}
The $\lambda$6300--8825 \AA$\;$wavelength regime contains many temperature-sensitive and surface gravity-sensitive
features diagnostic for classifying late-type stars.
To ensure that we could accurately classify our program stars, we observed a range 
of spectral main-sequence
standards (K5--L1.5), giant standards (K7--M9), and previously-identified USco 
objects (K3--M8;
 Preibisch \& Zinnecker 1999, Ardila et al. 2000, Mart\'{i}n et al. 2004).  In 
addition, we took
observations of spectroscopically confirmed 1-Myr Taurus members and 100-Myr 
Pleiades stars during
an observing run in December 2004 with the same telescope and 
instrument set-up. 
Together these observations provide a
large range in both temperature and surface gravity which we use for aid 
in classifying program stars. 

In this section we use molecular absorption features
to derive a spectral type for each object.  From atomic and molecular surface gravity diagnostics we 
determine which objects have low, PMS-type gravity and are therefore likely association members.  The effects 
of reddening and veiling are explored and considered in both temperature and 
gravity determination.

\subsection{Temperature Classification}
\begin{figure}
\epsscale{0.8}
\plotone{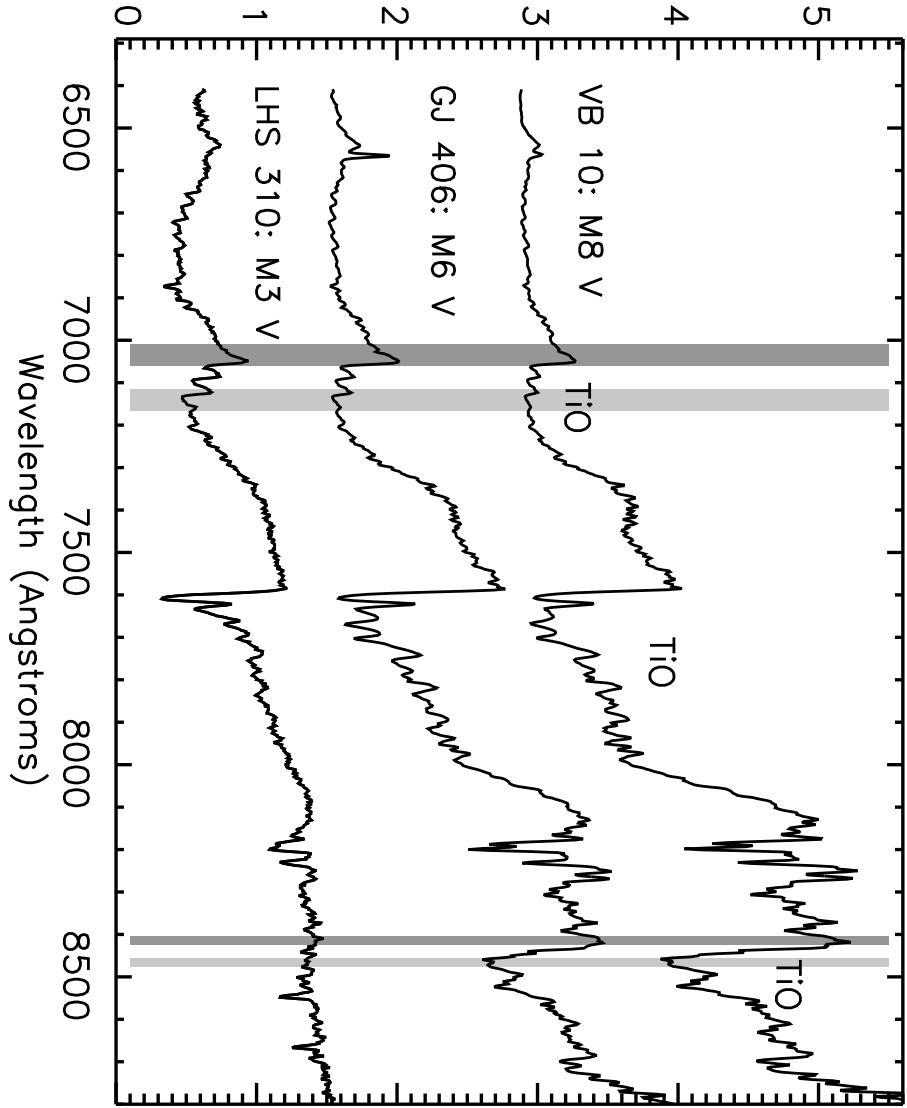}
\caption{Spectra for M3, M6, and M8 type dwarf stars.  Dominant TiO features are labeled.  
Light and dark shaded regions respectively show location of the TiO
and continuum bands used in spectral classification.}
\label{fig:tspec}
\end{figure}

Figure~\ref{fig:tspec} shows spectra for dwarf stars of spectral types M3, M6 \& M8.  
The dominant molecular absorption features present
are attributed to titianium-oxide (TiO), the strongest of which are labeled.  TiO 
absorption increases from mid-K thru $\sim$M7 spectral types at which point its 
strength begins to decrease and 
vandium oxide (VO) 
absorption starts to dominate spectra by early L types.  
We have defined or adopted from the literature several band indices to measure the 
strength 
of the TiO absorption features.  We find two in particular to be diagnostic. 
TiO-7140
 measures
the strength of TiO $\lambda$7140 \AA$\;$absorption compared to a 
continuum band at $\lambda$7035 \AA$\;$(Wilking et al. 2005).  This index is  
defined as TiO-7140=F$_{\lambda7035}$/F$_{\lambda7140}$ with bandwidths of 50\AA.
TiO-8465 is a new index defined by us (TiO-8465=F$_{\lambda8415}$/F$_{\lambda8465}$) to measure the strength of TiO $\lambda$8465 
\AA$\;$absorption 
compared to a continuum band at $\lambda$8415 \AA.
Bandwidths for the TiO-8465 index are 20 \AA.
In Figure~\ref{fig:tspec}, light and dark shaded regions respectively show 
location of the TiO and continuum bands used in our analysis.

\begin{figure}
\scalebox{0.4}{\includegraphics{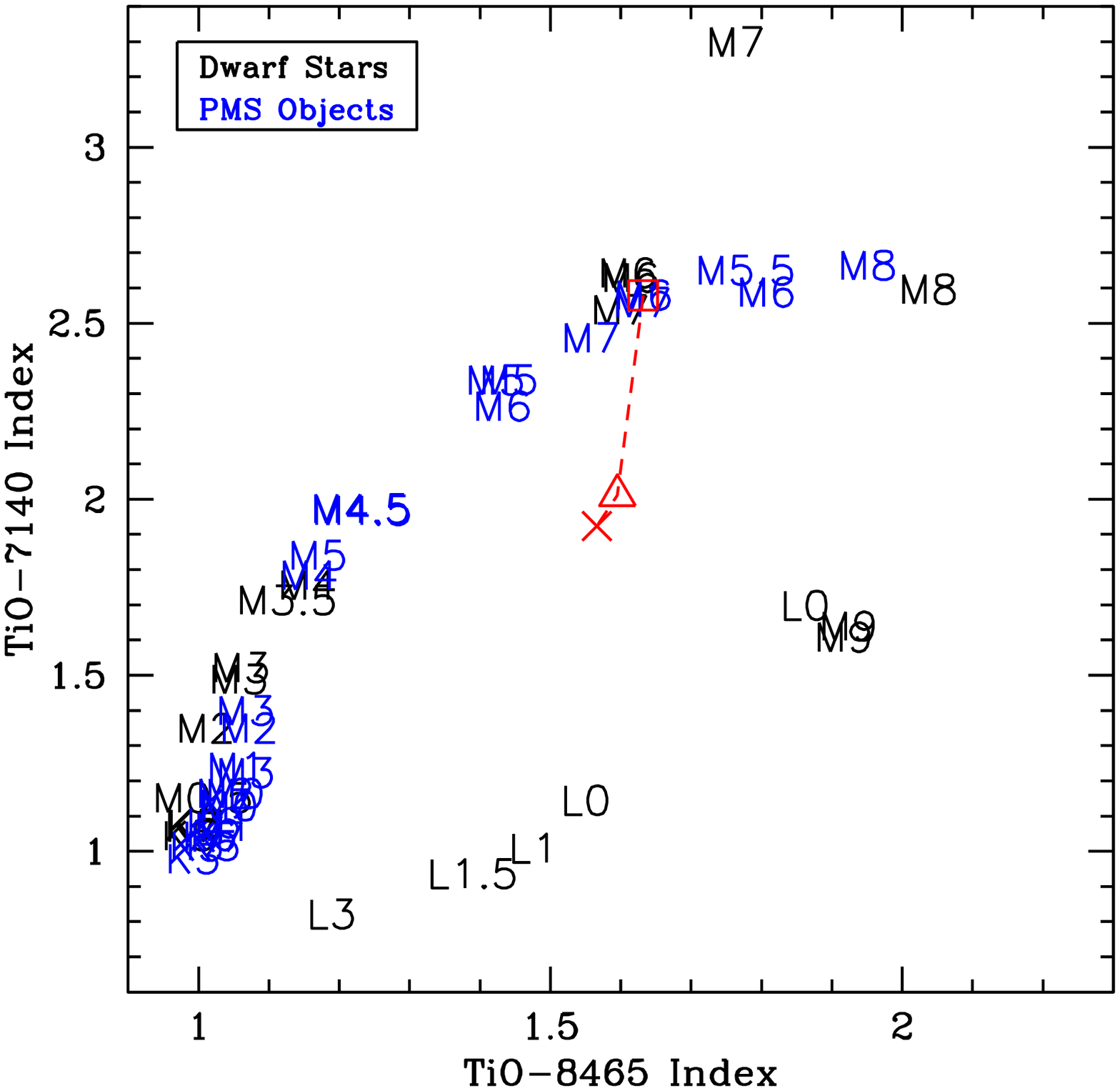}}
\hspace*{0.2in}
\scalebox{0.4}{\includegraphics{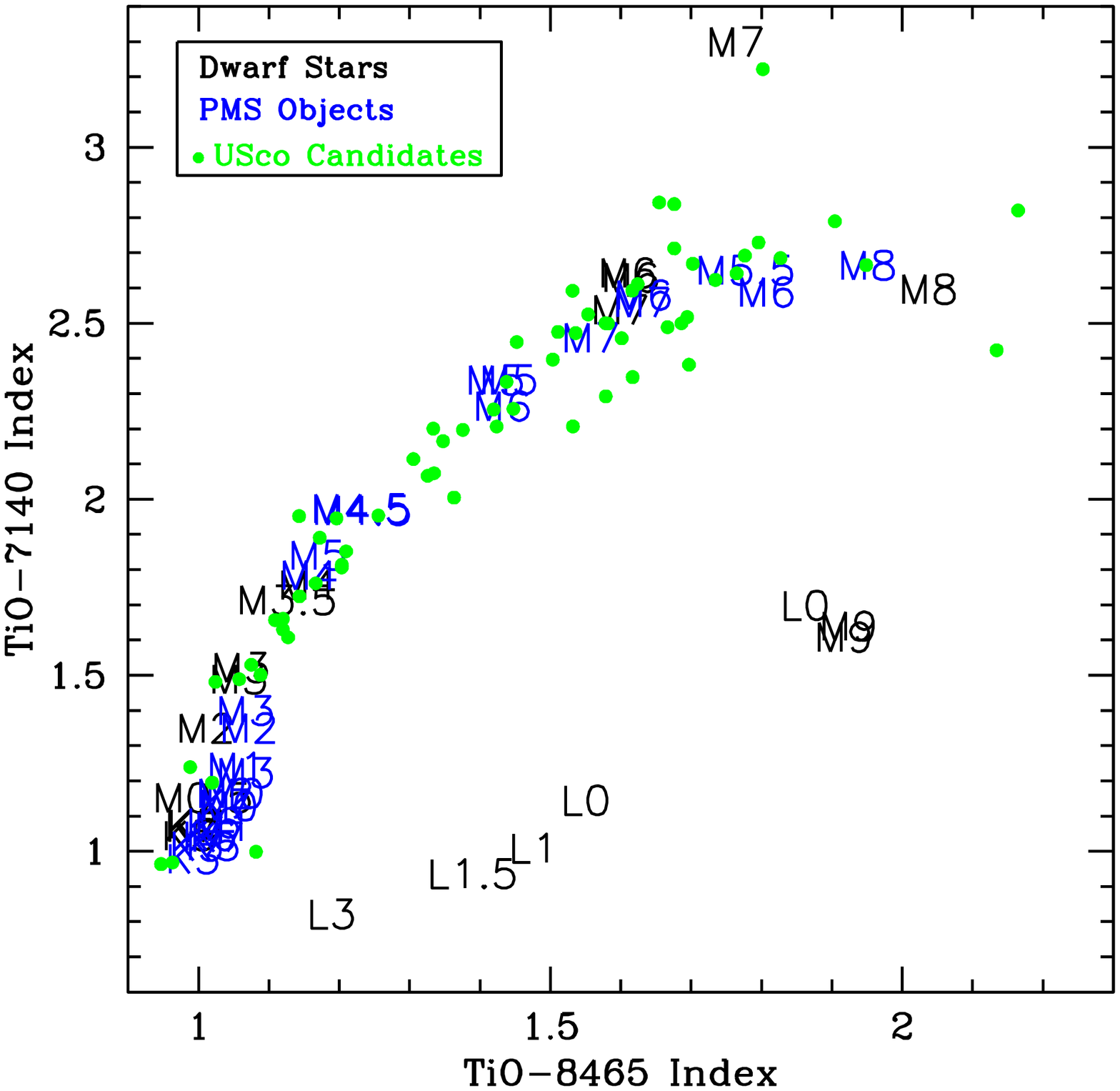}}
\caption{Left panel shows two
temperature-sensitive indices, TiO-7140 vs TiO-8165.  Spectral types represent 
measured indices for dwarf stars (black) and known USco and Taurus members (blue).  
The right panel adds green circles 
corresponding to measured indices for program stars.
In the left panel, connected red symbols indicate the effects of veiling 
from a cool circumstellar disk (box) or a hot accretion shock (triangle and `X') 
on measured indices for a single star (see text for more details).  
}
\label{fig:tind}
\end{figure}

The left panel of Figure~\ref{fig:tind} shows a plot
of TiO-8465 vs. TiO-7140;
black spectral types are
measured indices for field dwarf spectra and blue spectral types are measured 
indices for
known USco and Taurus members.  This diagram is particularly useful for 
classifying objects
with spectral types $\sim$M3--L3.  We find no surface gravity dependence between 
measured indices for the dwarf, USco, and Taurus standard objects.  
In the right panel of Figure~\ref{fig:tind}, green circles correspond to 
measured indices for programs stars.  As can be
seen, the program star measurements follow the locus laid out by the standard 
stars extremely well.  We find six outliers which sit below the main locus of data 
points.
In all six cases, the star is confirmed to exhibit low gravity signatures 
(Section 3.2) and we attribute the position in Figure~\ref{fig:tind} to a small 
amount of veiling 
present in its spectra (Section 3.3).
These objects are further discussed in Section 4.2.

Spectral types were determined first from 
quantitative 
analysis using an object's measured
TiO indices.  More weight was given to the value of the TiO-8465 index which we 
find particularly insensitive to the effects of reddening and veiling, as 
discussed in Section 3.3.
Using the TiO indices alone we can not
uniquely recover spectral types for the dwarf and PMS standards.  This ambiguity is due
in part to uncertainties in published spectral types and in part to 
systematic differences between spectral classification schemes of different authors.
Therefore, it was necessary to examine each spectrum by eye and use the information
from the entire spectral range in final type determination.  
All program spectra were compared visually to 
a large grid of standard spectra.  Typical revisions to the quantitative spectral
types were   
at the level of 1--2 subclasses.

\subsection{Surface Gravity Assessment}

\begin{figure}
\epsscale{0.8}
\plotone{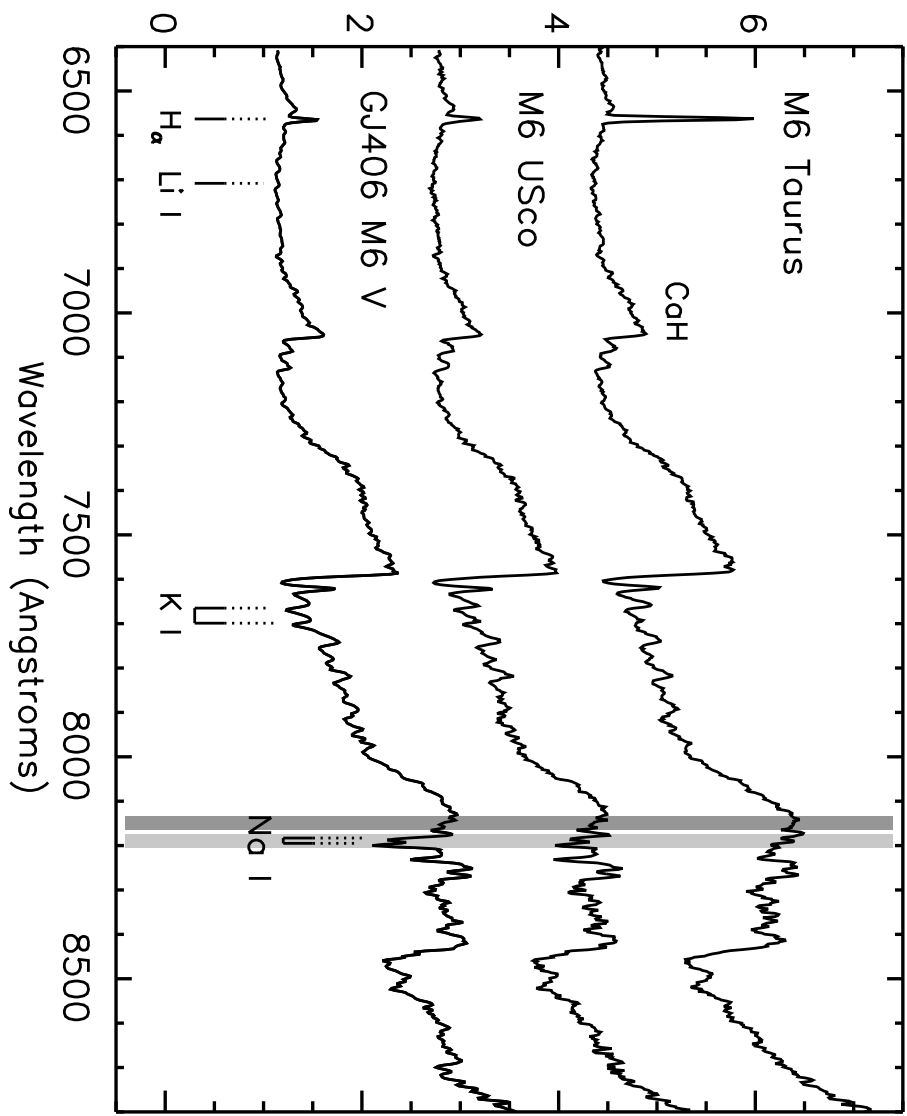}
\caption{Sequence of three M6 stars: a high-gravity main-sequence star (GJ406), 
a 5-Myr intermediate-gravity USco object (DENIS-P16007.5-181056.4) and a 1-Myr Taurus object (MH05).  Gravity-
sensitive absorption features are labeled and increase with increasing stellar age 
and gravity.  Light and dark shaded regions respectively show location of the Na
and continuum bands used in surface gravity analysis.}
\label{fig:gspec}
\end{figure}

Accurate spectral classification requires determining both the type (indicative
of temperature)
and the luminosity class (indicative of gravity) of our candidates.
Several robust surface gravity sensitive features exist in the $\lambda$6300--8800 
\AA$\;$ 
spectral regime
which can be assessed in low resolution spectra and used to distinguish high 
gravity dwarf stars from younger PMS objects still undergoing contraction.
The three most prominent features are due to NaI ($\lambda$8183 \& $\lambda$8195 
\AA) and KI 
($\lambda$7665 \& $\lambda$7699 \AA) lines and CaH molecular absorption at 
$\lambda\lambda$6975 \AA.
Figure~\ref{fig:gspec} shows a sequence of three M6 standard stars: a high-gravity 
main-sequence star (GJ406), a 5-Myr intermediate-gravity USco object (DENIS-P16007.5-181056.4) and
a 1-Myr lower-gravity Taurus object (MH05).  Gravity sensitive absorption features are 
labeled
and increase in strength with increasing stellar age and gravity.

We have developed a gravity-sensitive index, Na-8189, which measures the strength 
of the Na I
doublet at $\lambda$8189 \AA$\;$compared to the strength of a continuum band at 
$\lambda$8150 \AA.  Each band is 30 \AA$\;$wide.  The index is defined as
Na-8189=F$_{\lambda8189}$/F$_{\lambda8150}$.
The left panel of Figure~\ref{fig:gind} shows a plot of the temperature-sensitive 
TiO-7140 index vs the gravity-sensitive Na-8189 index.
Black and blue spectral types respectively represent measured indices for dwarfs 
and known PMS stars,
similar to Figure~\ref{fig:tind}.
Cyan spectral types show measured indices for low-gravity giant standards.  
We find the Na-8189 index to be a robust diagnostic that clearly separates low, 
intermediate and high 
gravity for spectral types later than M2.  Objects with spectral types 
earlier than $\sim$M2
do not exhibit substantial Na I absorption.  Measured Na-8189 indices 
for 100-Myr Pleiades stars are indistinguishable from those
for dwarfs at similar spectral types.  In the right panel of Figure~\ref{fig:gind} 
green circles correspond to 
measured indices for programs stars.  A large fraction, 43/62, of the candidate 
objects have measured Na-8189
indices consistent with their having surface gravity less than that of field 
dwarfs 
at similar spectral types. 

Gravity signatures in all spectra were verified visually.  For only 
one star did we find a visual assessment which disagreed with the quantitative 
result of the NA-8189 index.
SCH16224384-19510575 appears to be a giant star based on the 
Na-8189 index alone but was determined from visual inspection to be a young, PMS-type 
object based on the overall shape of its spectrum longward of $\sim$8200\AA$\;$ and strong H$\alpha$ emission (Section 3.4).  This object is discussed further in Section 4.2.

\subsection{Stellar and Interstellar Processes Which Can Affect Spectral 
Classification}

In this section we explore possible biases in the spectral classification due to
the effects of stellar and interstellar reddening, or spectral veiling produced 
either from a cool circumstellar disk or from a hot accretion shock.  We assess the effects 
these processes can have both on the quantitative TiO and Na indices and on the 
overall appearance of a spectrum.

\subsubsection{Extinction}

Hot OB stars have dispersed much of the dense gas and dust in the 
USco region; 
therefore, we do not 
expect individual members to show more than moderate extinction. We 
derived an approximate A$_V$ 
measurement for each star by visually dereddening its spectrum 
until its slope matched that of a standard star of the same spectral type.  We find
that 80\% of the newly identified members have A$_V <$1 based on spectral 
slope from $\lambda$6300--8825 \AA, and only three appear to have
A$_V$ between 2 and 3 mag.  More precise extinction values are derived from each object's 
spectral type and colors in Section 4.1,
and in all cases we find agreement to within one magnitude of our visual estimate.

To assess the effect reddening has on our quantitative classification indices we 
artificially reddened all standard stars by A$_V$=10 mag, the 
maximum extinction inferred from large-beam dust measurements towards our
survey region excluding the young $\rho$ Ophiuchus molecular cloud core
(Schlegel, Finkbeiner, \& Davis 1998).  We measured index strengths for artificially reddened spectra and find 
average index shifts
of -0.1
for TiO-7140 and -0.04 for TiO-8465.  
As can be seen from Figure~\ref{fig:tind}, index shifts at this level are not 
sufficient to affect our quantitative temperature
determination from TiO indices at the 0.5 subclass level.  Ten magnitudes of extinction could affect the Na-8189 
index by an average shift of -0.05. However at a level of reddening more consistent
with actual measured values for new members (A$_V \lesssim$2; Section 4.1, Table 2), we find a much more moderate shift of -0.01 which would not be sufficient to make a dwarf
star look like a PMS star for objects between M2 and M9.

\subsubsection{Optical and Near-infrared Veiling}

A young star spectrum can be veiled at UV/optical wavelengths due to excess 
emission 
from an accretion shock, and in the infrared due to thermal emission from dust 
grains
in a circumstellar disk.  
In both cases the excess emission veils (decreases) the strength of the molecular 
absorption features
used in classification and will cause a star to be systematically classified too 
hot (early) in spectral type.  The veiling index is defined as $r_\lambda = 
F_{\lambda ex} / F_{\lambda ph}$, where ``ex'' indicates excess and ``ph'' 
indicates photosphere.  
We tested several scenarios to determine the existence and 
magnitude of any bias in the spectral types attributable to veiling.

\begin{figure*}
\scalebox{0.4}{\includegraphics{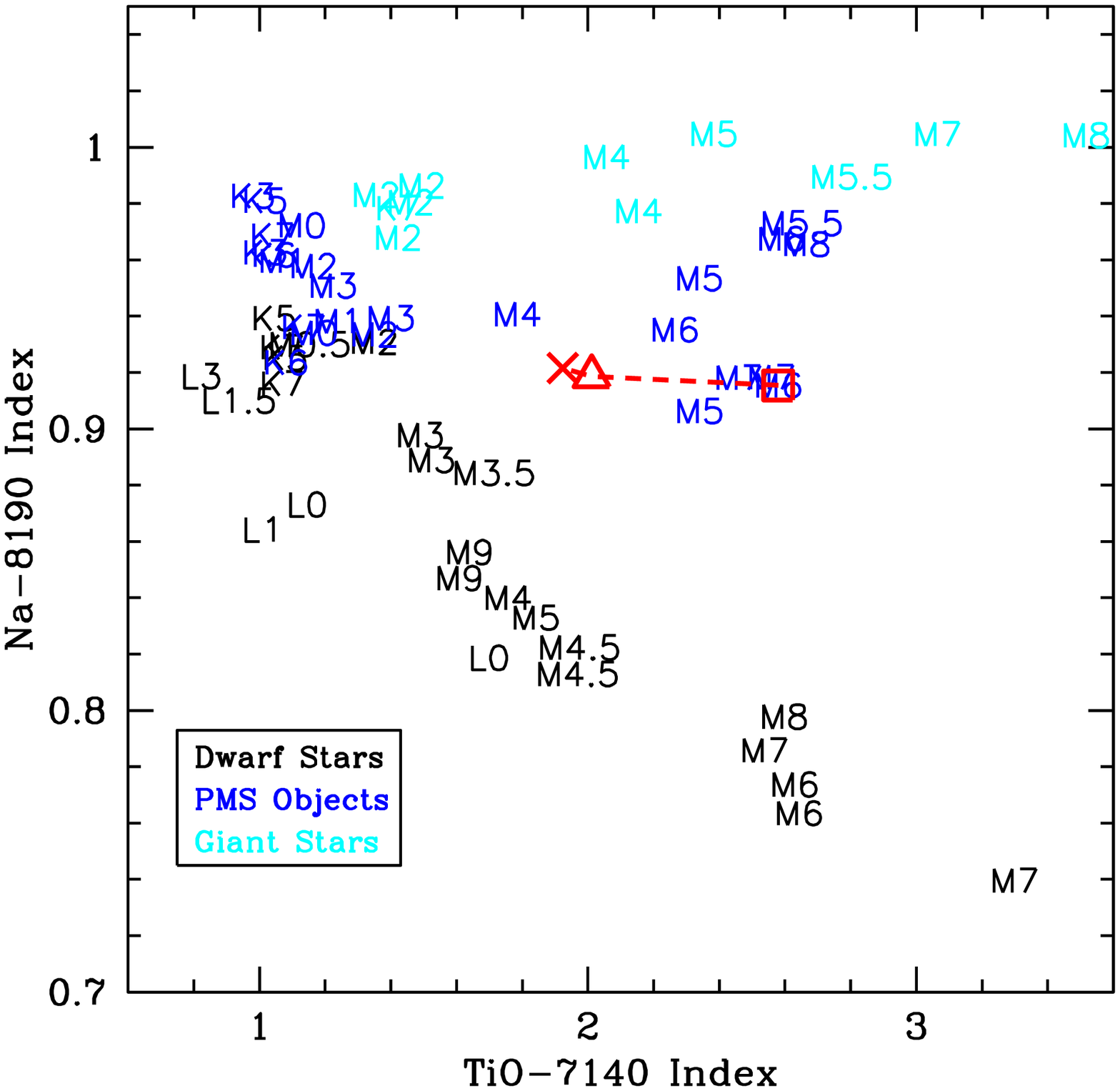}}
\vspace*{0.2in}
\scalebox{0.4}{\includegraphics{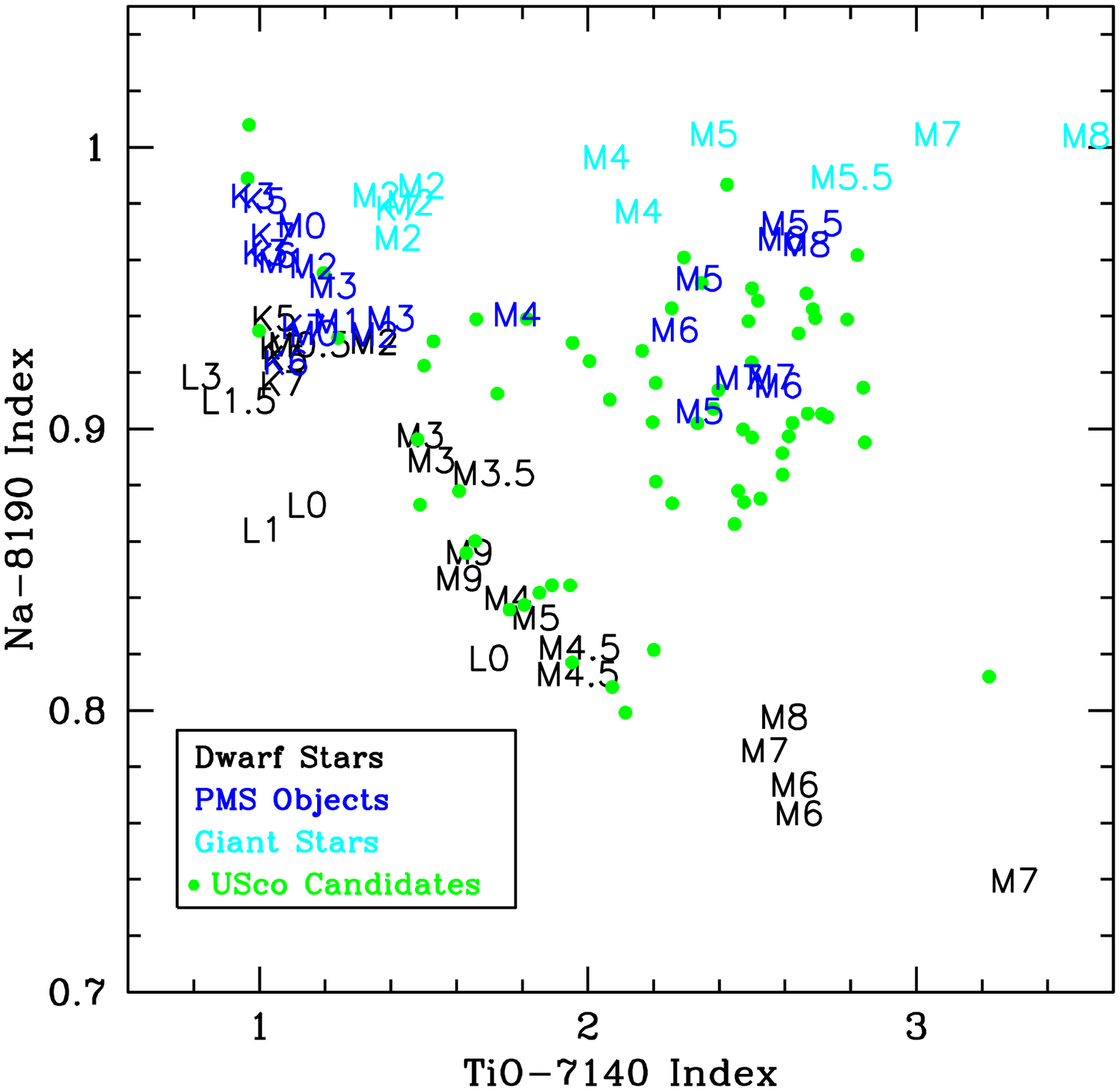}}
\caption{Same as Figure~\ref{fig:tind} but for the TiO-7140 index vs. the gravity-sensitive Na-8195 index.  Cyan spectral types represent measured indices for low gravity giant stars.}
\label{fig:gind}
\end{figure*}

First, we added to all spectra a T$_{eff}$=1400 K
blackbody consistent with a cool disk around a low mass star or brown dwarf at 
a veiling level of r$_K$=0.6, corresponding to the median near-infrared veiling 
value  
for K7-M0 classical T Tauri stars (Meyer et al. 1997).
The Wein tail of such a blackbody could affect our spectral indices around 0.8$\mu$m.
We remeasured indices for veiled standard 
spectra and found, as 
expected, excess thermal emission from a cool disk has very little effect on 
optical spectra.  
Results from veiling experiments are shown for a 
single star (DENIS-P16007.5-181056.4, M6 USco member; Mart\'{i}n et al. 2004) as 
connected red symbols in the left panels of 
Figure~\ref{fig:tind} and Figure~\ref{fig:gind}. 
We see similar results for all M-type (SpType $>$ M2) standards artificially 
veiled using these techniques. 
The re-derived indices for DENIS-P16007.5-181056.4 plus near-infrared veiling described
above are shown as red squares which lie practically on top of the original `M6' blue
points in both figures. 

Next, we investigated the effects of veiling from a hot accretion shock.
We used a value of r$_{6500}$=0.6, equivalent to the average optical
veiling
value for late K and early M stars (White \& Hillenbrand 2004).  We added a 
T$_{eff}$=8000 K blackbody at this level to all standard spectra and re-derived the 
spectral classification indices (triangle symbols Figures~\ref{fig:tind} \& ~\ref{fig:gind}).  We 
experimented by adding a continuum excess of constant flux ($F_{ex} = C$) 
at the 
r$_{6500}$=0.6 level ('X' symbol Figures~\ref{fig:tind} \& ~\ref{fig:gind}), shown by White \& 
Hillenbrand (2004) to be more consistent with observations than a hot blackbody.
In both cases, the strength of the TiO-7140 index decreases substantially while 
the change in the TiO-8465 and Na-8189 indices are much smaller.
For this reason, we rely primarily on the TiO-8465 index for temperature classification and believe our surface gravity assessment from the Na-8189 index to be robust 
to effects of veiling.

\subsection{Emission Lines}

\begin{figure}
\plotone{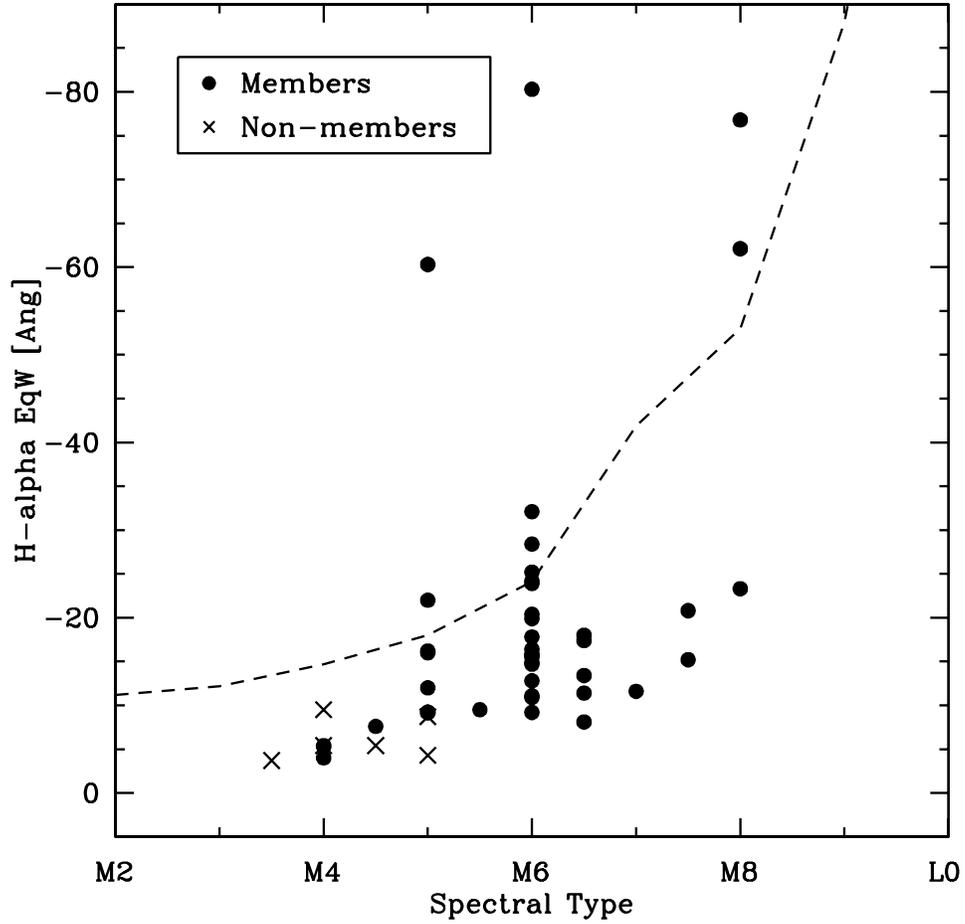}
\caption{Measured H$\alpha$ equivalent widths for all spectra with measurable H$\alpha$ as a function
of spectral type.  Circles represent objects confirmed to be new members; 'X's 
represent objects confirmed to be field dwarfs.  The dotted line is the empirical
accretor/non-accretor upper limit derived by Barrados y Navascu\'{e}s \& Mart\'{i}n (2003).} 
\label{fig:halp}
\end{figure}

The only prominent emission line observed in any of the spectra is H$\alpha$
which, seen in the spectra of young stars and brown dwarfs, is 
predominantly 
created via one of two mechanisms.
Weak, narrow H$\alpha$ lines are presumed to originate from active chromospheres 
whereas strong, broad and/or asymmetric lines can be produced from high-velocity, 
infalling accretion or strong winds.  
Barrados y Navascu\'{e}s \& Mart\`{i}n (2003) have proposed an empirical,
 spectral-type--H$\alpha$ equivalent width ($W$(H$_\alpha$))
relation to describe the upper limit of non-accreting stars and brown 
dwarfs based on the chromospheric saturation limit observed in young open 
clusters.
Figure~\ref{fig:halp} plots measured H$\alpha$ equivalent widths for all spectra 
as a function of spectral type, 
shown with the Barrados y Navascu\'{e}s \& Martin (2003) empirical accretor/non-
accretor division.
Four objects (SCH16222156-22173094: M5,  SCH16103876-18292353: M6, SCH16224384-19510575: M8, SCH16235158-23172740: M8; see Table 1) exhibit very strong emission ($W$(H$\alpha$)$<$-60 \AA) 
at levels substantially above those measured for the majority of sources in our sample
and are possibly still undergoing active accretion. High resolution spectroscopy is needed to assess further the evidence for accretion based on line profiles/shape. 

\subsection{Summary of Spectral Classification and Analysis}

We have determined spectral types and surface gravity estimates for 62 objects
towards the USco association.  These objects were selected as candidate PMS 
association members based on their optical and near-infrared colors and magnitudes.
In all cases, classification was done first using flux ratios of broad molecular 
absorption 
lines or narrow atomic lines to continuum levels.
Reddening and veiling were accounted for during the spectral classification
and gravity assessment processes.
All spectral types 
and surface gravity estimates were confirmed from visual comparison to standard 
spectra.  Unless otherwise noted in Table 1, spectral type errors are $\pm$0.5 
SpType.

Table 1 lists optical and near-infrared 2MASS photometry, measured spectral indices, 
spectral types and H$\alpha$ equivalent widths.  We also list a qualitative 
surface gravity type: `USco' and `dwarf' labels indicate a star
has surface gravity signatures consistent with those measured for known USco 
members or field dwarfs.  A value of `int' corresponds to the object having 
gravity signatures between those of known USco members and field dwarfs, 
indicating it is likely a more evolved association member, a member of one of the neighboring Sco-Cen subgroups, or a very young field
object.  As mentioned previously, we find Na-8189 indices for Pleiades
objects to be indistinguishable from those measured for dwarf stars.  Therefore
the `int' objects are most probably between 5 and $\sim$100 Myr. 
For the remainder of this paper we consider all objects with gravity classification 
`USco' or `int' as new association members.

\section{USco Population Analysis}

\subsection{HR-Diagram}

In this section we combine each new member's spectral type and photometry to derive 
values for its luminosity
and effective temperature and place it on a theoretical HR-diagram. 
As described in Section 2, the final calibration of Quest-2 optical data is still under revision.  Thus, because of the reliability and uniformity of the 2MASS
survey
we chose to use $J$-band magnitudes and $(J-H)$ colors to derive luminosities.
An empirical fit to BC$_J$ as a function of spectral type was determined from the observational 
data of Leggett et al. (1996, 2002; spectral types M1-M6.5 and M6-L3).
  We adopted intrinsic colors, extinction, and 
effective temperatures using the methods described in 
Slesnick, Hillenbrand \& Carpenter (2004).  Derived quantities are given in Table 
2.

In Figure~\ref{fig:hr} we present an HR-diagram for the 43 newly identified low 
mass members of USco, shown with the PMS model tracks and isochrones of 
 DM97.  The most commonly used PMS models for low mass stars and brown dwarfs are 
those derived by DM97 and Baraffe et al. (1998) which differ primarily in their 
atmospheric approximations and treatment of convection.
No models to date have consistently reproduced dynamical masses for young low mass 
objects 
(M$<$1.2 M$_\odot$; Hillenbrand \& White 2004).  We therefore do not attempt to 
derive 
masses and ages for new members at this time.
Both models suggest similar mass ranges for our data of 0.02M$_\odot$ $< M <$ 0.2M$_\odot$,
though predicted masses for individual objects can vary by up to 0.07M$_\odot$.
As illustrated in Figure~\ref{fig:hr}, we have identified a low
mass stellar population of age roughly consistant with the 5 Myr
age inferred in previous work on the intermediate mass (6M$_\odot$$<M<$0.1M$_\odot$) members of USco (Preibisch et al. 2002).

\subsection{Interesting Objects}

\begin{figure}
\epsscale{0.8}
\plotone{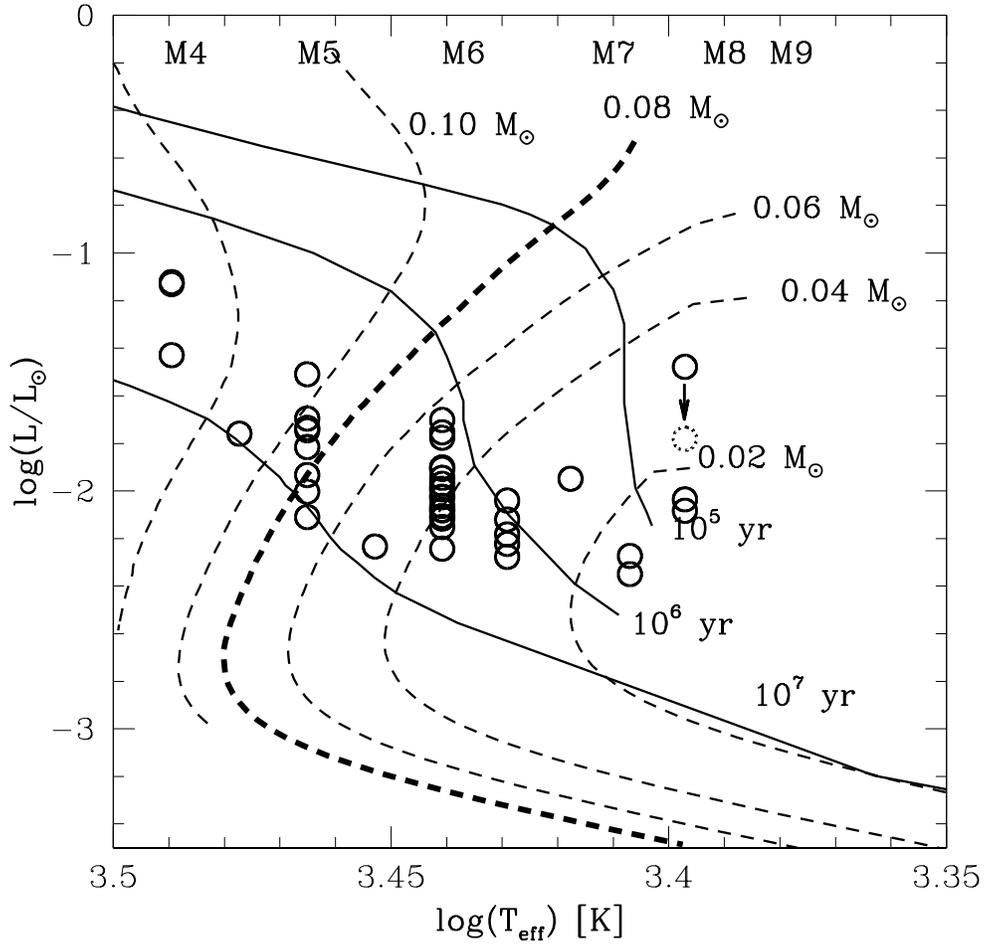}
\caption{HR diagram for new PMS objects found in this work, shown with model 
tracks and isochrones of DM97. The sample is consistant with an age of $\sim$5 Myr and contains 
masses spanning the brown dwarf to stellar regimes.}
\label{fig:hr}
\end{figure}

In this section we discuss objects and empirical observations worthy of comment.
The one outlier to the main locus of USco points in Figure~\ref{fig:hr} is object 
SCH16224384-19510575 which appears over-luminous and extremely young compared to the other sources.  
This object also has strong H$\alpha$ emission (see Section 3.4)
with a slightly asymmetric profile.  It is unlikely that this object is a 
single, extremely young ($<$100,000 yr based on
HR diagram placement) association member. 
The object could be a young, PMS-gravity 
foreground object that happens to fall within our line of sight.
The simplest explanation is that
SCH16224384-19510575 is an unresolved binary.  Assuming typical seeing at Palomar under photometric conditions of $\sim$1.2'', any pair with 
separations $\lesssim$175 AU would not be resolved in our data.  If we assume 
SCH16224384-19510575 
consists of 2 equal-luminosity objects, its placement in the HR-diagram becomes 
more consistent with the main locus of association members.  This effect is 
illustrated as an arrow plus dotted symbol in Figure~\ref{fig:hr}.

The six objects with measured TiO indices below the main locus of points (see Figure~\ref{fig:tind}) are
SCH16103876-18292353 (M6), SCH16202523-23160347 (M5.5), SCH16213591-23550341 (M6),
SCH16222156-22173094 (M5), SCH16224384-19510575 (M8), and SCH16235474-24383211 (M6).  Three of these objects, one of which is the possible binary discussed above, have very strong H$\alpha$ emission and are possible accretors.  Three lie very close (within 
$\sim$1 deg) to the young $\rho$Oph molecular cloud. However, because $\rho$Oph and USco
lie at approximately the same distance, if they were escaped, $\lesssim$1 Myr $\rho$Oph members
we would expect to see them exhibit systematically higher luminosities than USco members of similar spectral type.
Based on Figure~\ref{fig:hr}, this phenomenon is not observed.  All six have a near-infrared excess 
based on $J-H,H-K_S$ colors (Figure~\ref{fig:ccd}; Table 1) indicating the presence of an inner circumstellar disk. 
In every case we find
spectral shapes consistant with a small amount of veiling (r$_\lambda\sim$0.1--0.2) in their spectra.

\begin{figure}
\plotone{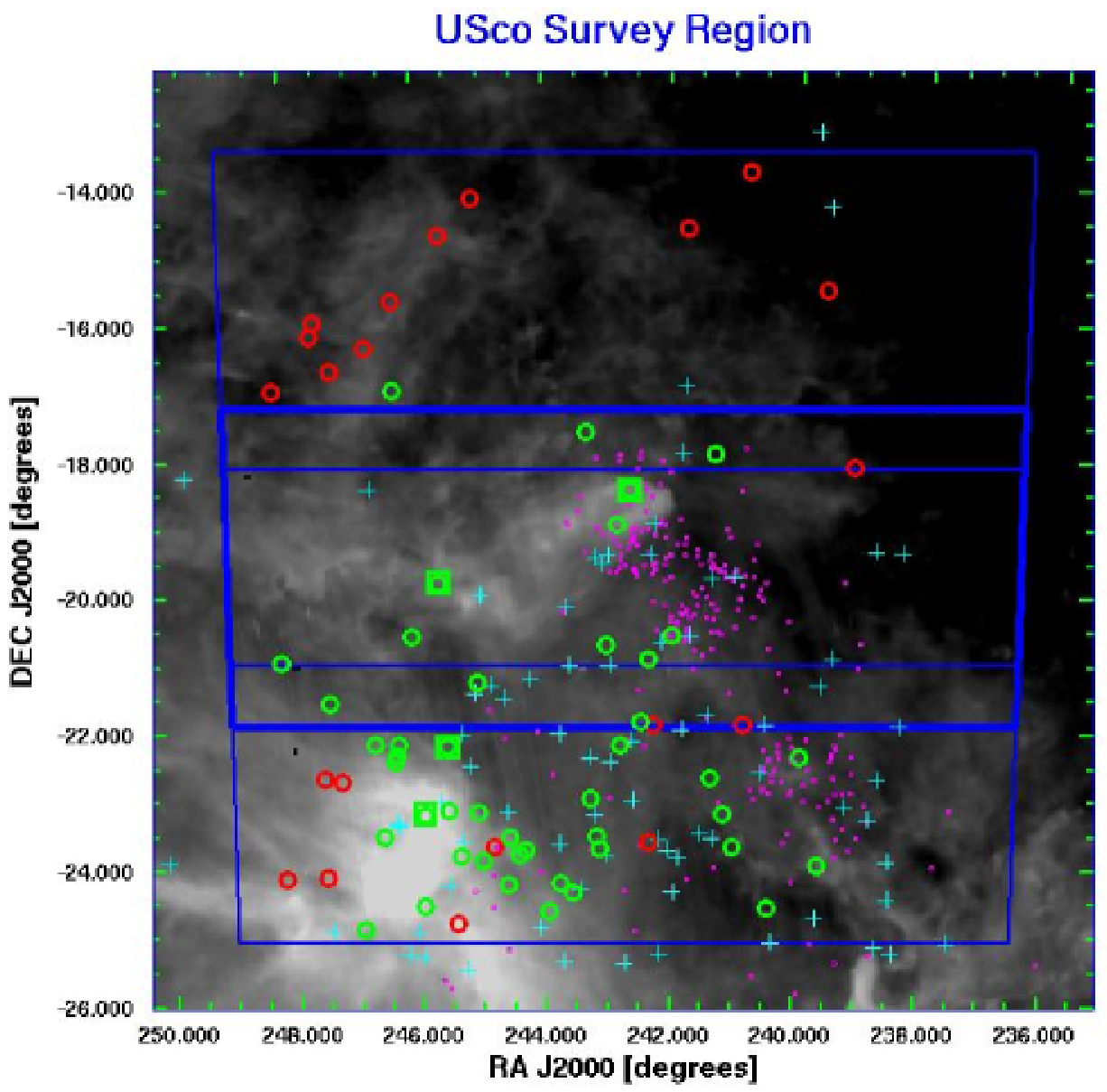}
\caption{Spatial distribution of the USco survey area outlined in blue, overlaid
on IRAS 100$\mu$m emission.  The thicker blue line denotes the outline of the monitoring scan region which was repeated 24 times.  New USco members identified from this work (43; green 
circles) are shown with previously known, spectroscopically confirmed low mass
members (196; Sp Type $\geq$K7 corresponding to M $\leq$0.6 M$_\odot$ at 5 Myr; small 
magenta circles) from the literature (see text), high mass Hipparcos members
(114; cyan pluses) and spectroscopic targets determined to be field
dwarfs (22; red circles).  New members which exhibit very strong H$\alpha$ emission 
are boxed.}
\label{fig:spat}
\end{figure}

The spatial distribution of the four strong H$\alpha$ emitters (likely accreting 
objects; boxed on Figure~\ref{fig:spat}) follows a ridge of dust outlined by 
100$\mu$m IRAS emission at the western edge of the association.  All four objects 
are 
located outside of previous surveys to search for low mass members, and outside 
most of the known high mass
association members (cyan pluses, Figure~\ref{fig:spat}). 
In general, we find some evidence for a trend of increased luminosity (within a 
given spectral type) for objects in the western portion of the cloud, independent of declination. 
Further spectroscopic observations will determine whether there exists a substantial population 
of young objects in these regions.  

\section{Summary and Future Work}

We have completed a large-area $R,I$ photometric survey in and near the Upper Scorpius 
region of recent star formation.
From these data we selected candidate new PMS association members
based on their optical and near-infrared colors and magnitudes.
We present here results from the first effort in our spectroscopic follow-up
campaign.  We observed 62 candidates and determined 43 (70\%) to be bona fide new
Upper Scorpius members. 
We derive an HR diagram for new members, mention noteworthy individual objects, and speculate on the spatial distribution 
of yet undiscovered low mass association members.

\begin{figure}
\plotone{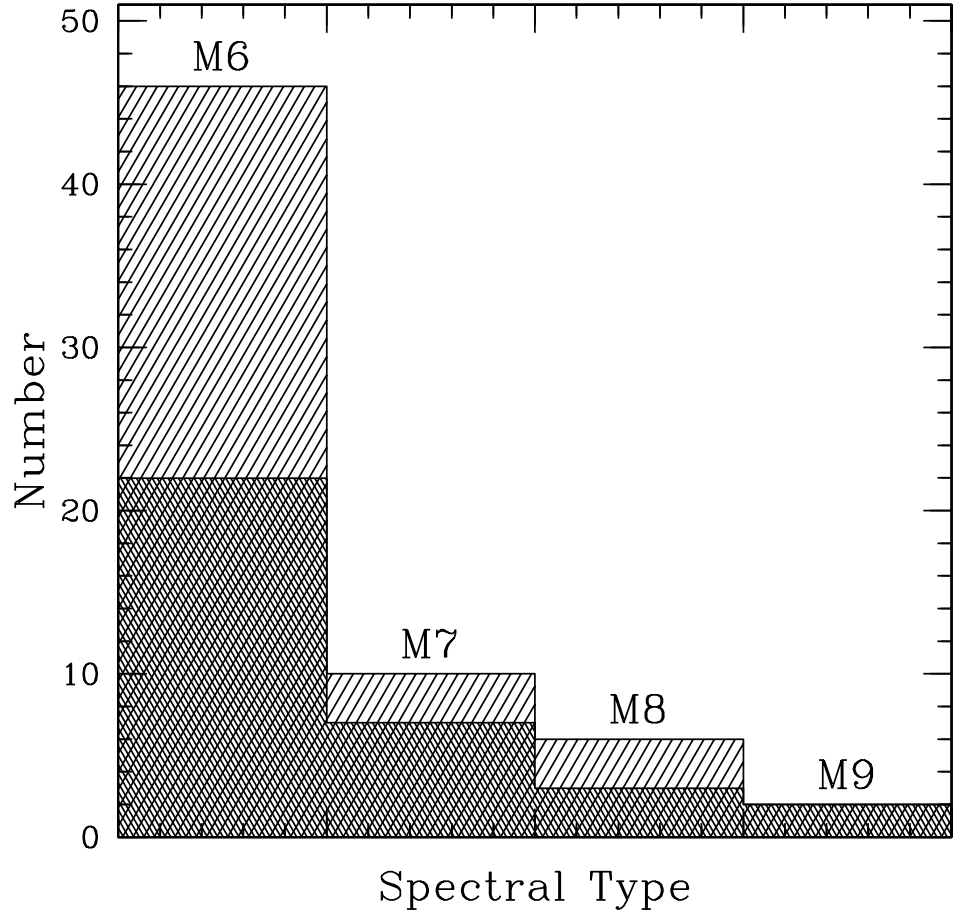}
\caption{Histogram of the number of brown dwarfs spectroscopically confirmed in
USco with the addition of our work (light shading) compared to a compilation of 
all previous results (dark shading).}
\label{fig:spechist}
\end{figure}

At an age of 5 Myr, all objects with spectral type $\geq$M6 are commonly 
considered
 to be substellar.  
Based on this criterion, from the 43 new members we identify 30 new brown dwarfs.
Prior to this work, 34 spectroscopically confirmed USco members had been 
identified 
at these spectral types (Preibisch et al. 1999, Ardila et al. 2000, Mart\`{i}n et 
al. 2004).  In Figure~\ref{fig:spechist} we present a histogram of the number of
brown dwarfs known with the addition of our work (light shading) compared to what 
was
known previously in the literature (dark shading).  As can be seen, with this 
study we have doubled the number of known substellar objects in Upper Scorpius.

In addition to the data presented here, we have taken spectra of $\sim$1000 
candidates with hydra/CTIO which will be presented in a forthcoming paper.
We have IRAC/$Spitzer$ data and approved
MIPS 24 $\mu$m/$Spitzer$ observations of 28 of the newly identified brown dwarfs 
which we analyze in Slesnick,
Carpenter, \& Hillenbrand (2006). 

\section{Acknowledgments}
The authors would like to thank the entire Quest-2 collaboration
and in particular David Rabinowitz, Anne Bauer, Jonathan Jerke and 
Adam Rengstorf for observing and processing the photometric driftscan data.
We are appreciative to Ashish Mahabal and Milan Bogosavljevic for their help
in understanding the Quest-2 systematics.  We thank
Russel White and David Ardila for discussions and insights which  
helped in our analysis.

\clearpage
\begin{deluxetable}{cccccccccccc}
\tabletypesize{\scriptsize}
\tablecolumns{12}
\tablewidth{0pc}
\tablenum{1}
\tablecaption{Measured Quantities for PMS Candidates with Observed Spectra}
\tablehead{
 \colhead{ID} & 
 \colhead{R} & 
 \colhead{I} & 
 \colhead{J\tablenotemark{c}} &
 \colhead{H\tablenotemark{c}} &
 \colhead{K$_S$\tablenotemark{c}} &
 \colhead{TiO-7140} &
 \colhead{TiO-8165} &
 \colhead{Na-8195} &
 \colhead{SpType\tablenotemark{d}} &
 \colhead{$W(H\alpha)$ [\AA]} &
 \colhead{Gravity\tablenotemark{e}}  
}
\startdata
SCH15563309-18074323 & 19.7 & 16.5 & 14.18 & 13.61 & 13.13 & 3.22 & 1.80 &  0.81 & M7 & -- & dwarf \\ 
SCH15582384-15310335 & 16.1 & 13.9 & 12.36 & 11.79 & 11.52 & 1.76 & 1.16 &  0.83 & M4.5 & -5.40 & dwarf \\ 
SCH15583162-24025411 & 17.5 & 15.0 & 13.16 & 12.51 & 12.20 & 1.95 & 1.25 &  0.93 & M4.5 & -7.60 & USco \\
SCH15594802-22271650 & 19.1 & 16.5 & 14.24 & 13.56 & 13.16 & 2.78 & 1.90 &  0.93 & M7.5 & -15.20 & USco \\
SCH16014768-24410152 & 18.5 & 15.9 & 13.87 & 13.27 & 13.00 & 2.44 & 1.45 &  0.86 & M5 & -16.00 & USco \\
SCH16031129-13454481 & 17.2 & 15.5 & 13.96 & 13.35 & 12.95 & 2.07 & 1.33 &  0.80 & M5 & -8.70 & dwarf \\
SCH16032871-21583609\tablenotemark{a} & 18.9 & 16.7 & 14.62 & 13.96 & 13.59 & 0.99 & 1.08 &  0.93 & K3: & -- & dwarf \\
SCH16040453-23463795 & 15.3 & 13.3 & 11.74 & 11.04 & 10.73 & 1.81 & 1.20 &  0.93 & M4 & -4.00 & int \\
SCH16044303-23182620\tablenotemark{b} & 19.0 & 15.9 & 13.81 & 13.19 & 12.85 & 2.72 & 1.79 &  0.90 & M6.5 & -18.00 & USco \\
SCH16051829-17562092 & 15.2 & 13.1 & 11.64 & 10.98 & 10.68 & 1.66 & 1.11 &  0.93 & M4 & -5.40 & int \\
SCH16053077-22462016 & 18.8 & 16.1 & 13.78 & 13.18 & 12.78 & 2.84 & 1.65 &  0.89 & M6 & -17.80 & USco \\
SCH16070403-14352587 & 15.9 & 14.2 & 12.28 & 11.73 & 11.41 & 1.89 & 1.17 &  0.84 & M4 & -- & dwarf \\
SCH16075850-20394890 & 18.4 & 15.8 & 13.59 & 12.95 & 12.58 & 2.47 & 1.53 &  0.89 & M6 & -14.90 & USco \\
SCH16091254-21582262 & 18.7 & 16.3 & 14.39 & 13.77 & 13.49 & 2.20 & 1.33 &  0.82 & M5 & -- & dwarf \\
SCH16092940-23431209 & 19.0 & 16.5 & 14.20 & 13.57 & 13.21 & 0.96 & 0.96 &  1.00  & $<$K3: & -- & dwarf \\
SCH16093018-20595409 & 18.8 & 16.2 & 13.99 & 13.35 & 12.98 & 2.59 & 1.53 &  0.89 & M6 & -11.10 & USco \\
SCH16095991-21554293 & 19.3 & 16.4 & 14.30 & 13.64 & 13.30 & 2.49 & 1.57 &  0.89 & M6.5 & -17.40 & USco \\
SCH16103876-18292353 & 19.2 & 16.5 & 13.96 & 13.16 & 12.64 & 2.29 & 1.57 &  0.96 & M6 & -80.30 & USco \\
SCH16111711-22171749 & 19.7 & 16.8 & 14.34 & 13.73 & 13.25 & 2.68 & 1.82 &  0.94 & M7.5 & -20.80 & USco \\
SCH16112959-19002921 & 19.0 & 16.2 & 13.67 & 12.90 & 12.44 & 2.49 & 1.68 &  0.94 & M6 & -20.40 & USco \\
SCH16121188-20472698 & 18.7 & 15.9 & 13.66 & 13.02 & 12.60 & 2.69 & 1.77 &  0.93 & M6.5 & -8.10 & USco \\
SCH16123758-23492340 & 18.5 & 16.0 & 13.93 & 13.28 & 12.91 & 2.47 & 1.51 &  0.87 & M6 & -15.80 & USco \\
SCH16124692-23384086 & 18.7 & 15.9 & 13.65 & 13.02 & 12.62 & 2.66 & 1.70 &  0.90 & M6 & -14.70 & USco \\
SCH16131212-23050329 & 19.1 & 16.4 & 14.05 & 13.44 & 13.00 & 2.62 & 1.73 &  0.90 & M6.5 & -13.40 & USco \\
SCH16132576-17373542 & 16.0 & 13.9 & 12.32 & 11.69 & 11.40 & 1.72 & 1.14 &  0.91 & M4 & -5.30 & int \\
SCH16141974-24284053 & 18.7 & 16.0 & 13.81 & 13.15 & 12.76 & 2.59 & 1.61 &  0.88 & M6 & -16.40 & USco \\
SCH16151115-24201556 & 19.0 & 16.3 & 14.23 & 13.58 & 13.17 & 2.33 & 1.43 &  0.90 & M6 & -10.90 & USco \\
SCH16155508-24443677 & 18.6 & 15.7 & 13.39 & 12.74 & 12.28 & 2.48 & 1.66 &  0.93 & M6 & -15.80 & USco \\
SCH16172504-23503799 & 18.8 & 16.1 & 13.74 & 13.01 & 12.63 & 2.25 & 1.44 &  0.87 & M5 & -12.00 & USco \\
SCH16174540-23533618 & 19.2 & 16.3 & 14.05 & 13.31 & 12.95 & 2.83 & 1.67 &  0.91 & M6 & -15.50 & USco \\
SCH16182501-23381068 & 18.9 & 16.1 & 13.72 & 12.88 & 12.44 & 2.06 & 1.32 &  0.91 & M5 & -9.20 & USco \\
SCH16183144-24195229 & 19.5 & 16.6 & 14.15 & 13.46 & 12.97 & 2.45 & 1.60 &  0.87 & M6.5 & -11.40 & int \\
SCH16192530-23470717 & 18.5 & 16.1 & 14.01 & 13.17 & 12.75 & 1.60 & 1.12 &  0.87 & M4 & -- & dwarf \\
SCH16200756-23591522 & 18.5 & 15.6 & 13.21 & 12.48 & 12.05 & 2.49 & 1.58 &  0.92 & M6 & -24.20 & USco \\
SCH16202127-21202923 & 18.5 & 15.5 & 13.39 & 12.74 & 12.40 & 2.61 & 1.62 &  0.89 & M6 & -23.90 & USco \\
SCH16202523-23160347 & 18.9 & 16.6 & 14.37 & 13.68 & 13.23 & 2.20 & 1.53 &  0.88 & M5.5 & -9.50 & USco \\
SCH16202753-14082840 & 16.0 & 13.8 & 12.22 & 11.56 & 11.27 & 1.85 & 1.20 &  0.84 & M4.5 & -- & dwarf \\
SCH16213591-23550341 & 19.5 & 16.5 & 13.94 & 13.19 & 12.73 & 2.38 & 1.69 &  0.90 & M6 & -19.90 & USco \\
SCH16214806-24542504 & 19.7 & 17.4 & 15.19 & 14.22 & 13.61 & 1.23 & 0.98 &  0.93 & M2 & -- & dwarf \\
SCH16221577-23134936 & 18.3 & 15.8 & 13.71 & 13.14 & 12.80 & 2.52 & 1.55 &  0.87 & M6 & -9.20 & int \\
SCH16222156-22173094 & 18.0 & 15.5 & 13.74 & 13.09 & 12.61 & 2.00 & 1.36 &  0.92 & M5 & -60.30 & USco \\
SCH16223315-14422746 & 17.6 & 15.3 & 13.49 & 12.84 & 12.51 & 2.11 & 1.30 &  0.79 & M5 & -- & dwarf \\
SCH16224384-19510575 & 17.0 & 15.0 & 12.35 & 11.61 & 11.15 & 2.42 & 2.13 &  0.98 & M8 & -62.10 & USco \\
SCH16235158-23172740 & 19.3 & 16.1 & 13.55 & 12.89 & 12.41 & 2.82 & 2.16 &  0.96 & M8 & -76.80 & USco \\
SCH16235474-24383211 & 19.1 & 16.0 & 13.31 & 12.49 & 11.92 & 2.34 & 1.61 &  0.95 & M6 & -12.80 & USco \\
SCH16242880-20385513 & 18.9 & 16.7 & 14.58 & 13.77 & 13.31 & 1.50 & 1.08 &  0.92 & M3 & -- & dwarf \\
SCH16252609-15401969 & 17.4 & 15.3 & 13.67 & 12.94 & 12.52 & 1.52 & 1.07 &  0.93 & M3 & -- & dwarf \\
SCH16252862-16585055 & 19.2 & 16.2 & 13.67 & 13.01 & 12.62 & 2.66 & 1.94 &  0.94 & M8 & -23.30 & USco \\
SCH16252968-22145448 & 18.0 & 15.4 & 13.19 & 12.49 & 12.11 & 2.25 & 1.41 &  0.94 & M5 & -16.20 & USco \\
SCH16253671-22242887 & 18.7 & 15.9 & 13.53 & 12.83 & 12.45 & 2.64 & 1.76 &  0.93 & M7 & -11.60 & USco \\
SCH16254319-22300300 & 16.8 & 15.0 & 13.02 & 12.40 & 12.09 & 2.19 & 1.37 &  0.90 & M5 & -9.20 & USco \\
SCH16263026-23365552 & 18.9 & 16.5 & 13.75 & 12.82 & 12.21 & 2.51 & 1.69 &  0.94 & M6 & -32.10 & USco \\
SCH16265619-22135224 & 18.5 & 15.7 & 13.48 & 12.83 & 12.41 & 2.71 & 1.67 &  0.90 & M6 & -28.40 & USco \\
SCH16270959-16204810 & 17.4 & 15.4 & 13.55 & 12.72 & 12.28 & 1.19 & 1.01 &  0.95 & M2 & -- & dwarf \\
SCH16274801-24571371 & 19.2 & 16.2 & 13.54 & 12.65 & 12.11 & 2.20 & 1.42 &  0.91 & M5 & -22.00 & USco \\
SCH16290665-22464968 & 18.6 & 16.2 & 14.03 & 13.19 & 12.77 & 1.65 & 1.10 &  0.86 & M4 & -9.50 & dwarf \\
SCH16291911-16410107 & 16.8 & 14.6 & 12.85 & 12.12 & 11.76 & 1.62 & 1.11 &  0.85 & M3.5 & -3.70 & dwarf \\
SCH16294877-21370914 & 16.9 & 14.5 & 12.52 & 11.86 & 11.52 & 2.16 & 1.34 &  0.92 & M5 & -9.20 & USco \\
SCH16300911-24111718\tablenotemark{a} & 16.1 & 14.7 & 13.38 & 12.84 & 12.56 & 1.48 & 1.05 &  0.87 & M2 & -- & dwarf \\
SCH16301483-22435159 & 18.8 & 16.7 & 14.57 & 13.71 & 13.20 & 1.48 & 1.02 &  0.89 & M3 & -- & dwarf \\
SCH16301682-15574807 & 18.0 & 15.6 & 13.59 & 12.86 & 12.53 & 1.95 & 1.14 &  0.81 & M5 & -4.30 & dwarf \\
SCH16303166-16093250 & 17.5 & 15.2 & 13.38 & 12.69 & 12.37 & 1.80 & 1.20 &  0.83 & M4 & -5.40 & dwarf \\
SCH16324726-20593771 & 17.9 & 15.5 & 13.45 & 12.85 & 12.47 & 2.39 & 1.50 &  0.91 & M6 & -25.20 & USco \\
SCH16325276-24110013\tablenotemark{a} & 17.1 & 15.9 & 14.33 & 13.69 & 13.40 & 0.96 & 0.94 &  0.98 & $<$K3 & -- & dwarf \\
SCH16325602-16582835 & 15.9 & 13.7 & 12.26 & 11.76 & 11.44 & 1.94 & 1.19 &  0.84 & M4 & -- & dwarf \\
\enddata
\tablenotetext{a}{Photometry for three targets changed significantly after spectroscopic observations took place, such that they would no longer have been chosen as photometric candidates.}
\tablenotetext{b}{The only object amoung our spectral targets previously identified in the literature
is SCH16044303-23182620 corresponding to UScoCTIO110 (Ardila et al. 2000).}
\tablenotetext{c}{Near-infrared photometry taken from 2MASS.}
\tablenotetext{d}{Spectral type errors are $\pm$0.5 for M subclasses. A `:' indicates a spectral type is less certain.}
\tablenotetext{e}{Qualitative 
surface gravity type `USco' and `dwarf' labels indicate a star
has surface gravity signatures consistent with those measured for known USco 
members or field dwarfs.  A value of `int' corresponds to the object having 
gravity signatures between those of known USco members and field dwarfs.}

\end{deluxetable}

\clearpage
\begin{deluxetable}{ccccc}
\tablewidth{0pc}   
\tablecolumns{5}
\tablenum{2}
\tablecaption{Derived Quantities for New USco Members}
\tablehead{
  \colhead{ID} & 
  \colhead{M$_J$} &
  \colhead{A$_V$} & 
  \colhead{log($T_{eff}$/K)} & 
  \colhead{Log ($L/L_\odot$)} 
}
\startdata
SCH15583162-24025411 & 7.18 & 0.66 & 3.48 & -1.76 \\ 
SCH15594802-22271650 & 8.35 & 0.33 & 3.41 & -2.27 \\ 
SCH16014768-24410152 & 8.03 & 0.15 & 3.47 & -2.11 \\ 
SCH16040453-23463795 & 5.64 & 1.14 & 3.49 & -1.13 \\ 
SCH16044303-23182620 & 7.99 & 0.05 & 3.43 & -2.12 \\ 
SCH16051829-17562092 & 5.62 & 0.85 & 3.49 & -1.12 \\ 
SCH16053077-22462016 & 7.97 & 0.04 & 3.44 & -2.11 \\ 
SCH16075850-20394890 & 7.69 & 0.38 & 3.44 & -1.99 \\ 
 SCH1609301820595409 & 8.09 & 0.39 & 3.44 & -2.15 \\ 
SCH16095991-21554293 & 8.38 & 0.43 & 3.43 & -2.28 \\ 
SCH16103876-18292353 & 7.69 & 1.78 & 3.44 & -1.99 \\ 
SCH16111711-22171749 & 8.54 & 0.00 & 3.41 & -2.35 \\ 
SCH16112959-19002921 & 7.46 & 1.52 & 3.44 & -1.90 \\ 
SCH16121188-20472698 & 7.79 & 0.24 & 3.43 & -2.04 \\ 
SCH16123758-23492340 & 8.00 & 0.48 & 3.44 & -2.12 \\ 
SCH16124692-23384086 & 7.77 & 0.29 & 3.44 & -2.02 \\ 
SCH16131212-23050329 & 8.25 & 0.00 & 3.43 & -2.22 \\ 
SCH16132576-17373542 & 6.38 & 0.52 & 3.49 & -1.43 \\ 
SCH16141974-24284053 & 7.87 & 0.54 & 3.44 & -2.06 \\ 
SCH16151115--24201556 & 8.32 & 0.42 & 3.44 & -2.24 \\ 
SCH16155508-24443677 & 7.46 & 0.49 & 3.44 & -1.90 \\ 
SCH16172504-23503799 & 7.59 & 1.31 & 3.47 & -1.93 \\ 
SCH16174540-23533618 & 7.91 & 1.26 & 3.44 & -2.08 \\ 
SCH16182501-23381068 & 7.29 & 2.37 & 3.47 & -1.82 \\ 
SCH16183144-24195229 & 8.15 & 0.76 & 3.43 & -2.18 \\ 
SCH16200756-23591522 & 7.09 & 1.18 & 3.44 & -1.75 \\ 
SCH16202127-21202923 & 7.48 & 0.41 & 3.44 & -1.91 \\ 
SCH16202523-23160347 & 8.31 & 0.95 & 3.45 & -2.23 \\ 
SCH16213591-23550341 & 7.78 & 1.38 & 3.44 & -2.03 \\ 
SCH16221577-23134936 & 7.91 & 0.00 & 3.44 & -2.08 \\ 
SCH16222156-22173094 & 7.76 & 0.66 & 3.47 & -2.00 \\ 
SCH16224384-19510575 & 6.35 & 0.75 & 3.40 & -1.48 \\ 
SCH16235158-23172740 & 7.75 & 0.00 & 3.40 & -2.04 \\ 
SCH16235474-24383211 & 6.96 & 2.07 & 3.44 & -1.70 \\ 
SCH16252862-16585055 & 7.86 & 0.02 & 3.40 & -2.08 \\ 
SCH16252968-22145448 & 7.09 & 1.14 & 3.47 & -1.73 \\ 
SCH16253671-22242887 & 7.55 & 0.67 & 3.42 & -1.95 \\ 
SCH16254319-22300300 & 7.12 & 0.38 & 3.47 & -1.74 \\ 
SCH16263026-23365552 & 7.15 & 3.02 & 3.44 & -1.78 \\ 
SCH16265619-22135224 & 7.57 & 0.39 & 3.44 & -1.94 \\ 
SCH16274801-24571371 & 6.99 & 2.82 & 3.47 & -1.70 \\ 
SCH16294877-21370914 & 6.53 & 0.73 & 3.47 & -1.51 \\ 
SCH16324726-20593771 & 7.64 & 0.06 & 3.44 & -1.97 \\ 
\enddata

\end{deluxetable}

\end{document}